\documentstyle[12pt,axodraw,epsfig]{article}
\def\mathrm#1{\mbox{\rm #1}}
\def\epsi{``$\epsilon$-approximation''~}
\def\Hb{\mbox{$H^0$~}}
\def\hb{\mbox{$h^0$~}}
\def\Ab{\mbox{$A^0$~}}
\def\MH{\mbox{$M_H$~}}
\def\Mh{\mbox{$M_h$~}}
\def\MA{\mbox{$M_A$~}}
\def\mH{\mbox{$m_H$~}}
\def\mh{\mbox{$m_h$~}}
\def\mA{\mbox{$m_A$~}}
\def\aeff{\mbox{$\alpha_{eff}$~}}
\def\tb{\mbox{$\tan\beta$~}}
\def\ra{\rightarrow}
\def\dmh{\mbox{$\delta M_h^{EPA}$~}}

\def\dszh{\mbox{$\delta\sigma_{Zh}^{EPA}$~}}
\def\dsah{\mbox{$\delta\sigma_{Ah}^{EPA}$~}}
\def\sqrts{\mbox{$\surd{s}$~}}
\def\admh{\mbox{$|\delta M_h^{EPA}|$~}}
\def\admH{\mbox{$|\delta M_H^{EPA}|$~}}

\def\adszh{\mbox{$|\delta\sigma_{Zh}^{EPA}|$~}}
\def\adsah{\mbox{$|\delta\sigma_{Ah}^{EPA}|$~}}

\begin{document}

\thispagestyle{empty}

\begin{titlepage}
\date{December 1995}
\title{
{\normalsize \hfill  hep-ph/9512441\\
\hfill KA-TP-16-1995}\\[1cm]
Production of neutral MSSM Higgs bosons in $e^+e^-$ collisions:
a complete 1-loop calculation.
}
\author{V. Driesen, W. Hollik,  J. Rosiek\thanks{on
leave of absence from Institute of Theoretical Physics, Warsaw University}
\and
Institut f\"ur Theoretische Physik\\
Universit\"at Karlsruhe, D-76128 Karlsruhe, Germany}
\maketitle
\begin{abstract}
We present the first complete
1-loop diagrammatic calculation of the cross sections for the neutral
Higgs production processes
$e^+e^-\ra Z^0h^0$ and $e^+e^-\ra A^0h^0$ in the minimal supersymmetric
standard model.  We compare the
results from the diagrammatic calculation with the corresponding
ones of the simpler and compact effective potential approximation and discuss
the typical size of the differences.
\end{abstract}
\end{titlepage}

\setcounter{page}{1}

\section{Introduction}
\label{sec:intro}

The Minimal Supersymmetric Standard Model (MSSM) predicts at least one
light neutral scalar Higgs particle with mass below $\sim 150$ GeV.
In order to experimentally detect possible signals of Higgs bosons and to trace
back as far as possible the physical origin of a produced scalar particle,
detailed studies for the decay and production processes of Higgs boson are
required.

As has been discovered several years ago~[1-3],
radiative corrections in the MSSM Higgs sector are large and have to be
taken into account for phenomenological studies. Several methods have been
developed to calculate the radiative corrections to the MSSM Higgs
boson masses, production and decay rates. Three main approaches have
been used for the calculation of 1-loop corrections:
\begin{itemize}
\item[a)] The Effective Potential Approach (EPA)~\cite{EPA}: This method
computes the dominant correction to the Higgs boson mass spectrum and
coupling constants in a simple and fast way and is thus most suitable for
numerical calculations, e.g. in Monte Carlo studies. In its practical
realizations, it is based on the following approximations:
\begin{itemize}
\item[--] only the corrections to the 2-point Green's functions are
included, and only the contributions from top/stop and bottom/sbottom
are taken into account
\item[--] the Green's functions are evaluated at momentum $p^2=0$, thus
neglecting the momentum dependence.
\end{itemize}
The EPA leads to expressions for the cross sections and decay branching
ratios in an effective Born approximation, where the tree level
quantities are replaced by the corrected Higgs boson masses and mixing angles.
\item[b)] The Renormalization Group approach (RGE)~\cite{RGE}.
This method also leads
to an effective Born approximation in the formulae for cross sections.
The 1-loop corrected Higgs boson masses and couplings are obtained
by using RG equations. Large logarithmic terms can be resummed, but realistic
mass spectra are not covered by this method since it relies on an effective
SUSY scale. The momentum dependence of the self-energies and 3- and
4-point functions are neglected also in this approach.
\item[c)] The diagrammatic calculation (Feynman Diagram Calculation,
FDC)~\cite{MYNPB,BRI}: The masses are calculated from the pole positions of
the Higgs propagators, and the cross sections are obtained from the full set
of 1-loop diagrams contributing to the amplitudes.
The version of ref.~\cite{MYNPB} is based on the on-shell renormalization
scheme. It takes into account:
\begin{itemize}
\item[--] the most general form of the MSSM lagrangian with soft
breaking terms,
\item[--] the virtual contributions from all the particles of the MSSM
spectrum,
\item[--] all 2-, 3- and 4-point Green's functions for a given process with
Higgs particles,
\item[--] the momentum dependence of the Green's functions,
\item[--] the leading reducible diagrams of higher orders corrections.
\end{itemize}
This method is technically most complicated, but also most accurate at
the 1-loop level and can be used as the reference frame for simpler
methods. Moreover, it allows to study the full parameter dependence of
cross sections and decay rates.
\end{itemize}

The experimental searches for Higgs bosons at LEP1~\cite{lepexp} and studies
for the future searches at higher energies~\cite{highen} conventionally make
use of the most compact
effective potential approximation. The phenomenological
study in ref.~\cite{JRAS} for LEP2 and higher energies, was done in the FDC,
but still not complete at 1-loop order. A check of the quality of the simpler
approximations and estimates
of their reliability in various energy and parameter ranges have also not yet
been performed so far.

In this paper we present the first complete 1-loop diagrammatic results for
the cross sections for the neutral Higgs production processes
$e^+e^-\ra Z^0h^0$ and $e^+e^-\ra A^0h^0$. In addition, we compare the
results from the complete diagrammatic calculation with the corresponding
ones of the simpler and compact effective potential approximation and discuss
the typical size of the differences.

Recently some papers on the leading 2-loop corrections
to the CP-even MSSM Higgs boson masses have been published~\cite{twoloop}.
The main conclusion is that 2-loop corrections are also significant and
tend to compensate partially the effects of 1-loop corrections.
The calculations are based on the EPA and RG methods. Since one of the main
emphasis of our study is to figure out the difference between complete
and approximate results in a given order, we have not implemented the 2-loop
terms. They would improve the 1-loop FDC results in the same way as the
approximations and thus do not influence the remaining differences which can
only be obtained by an explicit diagrammatic calculation.

In Section~\ref{sec:outline} we give a short description of the
computational schemes for the
masses and cross sections. This is followed in Section~\ref{sec:results}
by a discussion of the numerical results and a comparison of the various
approximations.

\section{Outline of the calculations}
\label{sec:outline}

The tree level potential for the neutral MSSM Higgs bosons can be
written as:
\begin{eqnarray}
V^{(0)} = m_1^2 H_1^2 +  m_2^2 H_2^2
+ \epsilon_{ij}(m_{12}^2 H_1^i H_2^j + H.c.)
+ {1\over 8} (g^2 + {g'}^2)(H_1^2-H_2^2)^2 + {1\over 4}g^2(H_1 H_2)^2
\label{eq:pot0}
\end{eqnarray}
Diagonalization of the mass matrices for the CP-even and the CP-odd
scalars, following from the potential~(\ref{eq:pot0}), leads to three
physical particles: two CP-even Higgs bosons
$H^0$,\hb and one CP-odd
Higgs boson \Ab, and defines their tree-level masses \mH, \mh and \mA, with
$\mH>\mh$,  and the mixing angles $\alpha$, $\beta$.

\bigskip

The 1-loop radiative corrections significantly modify  the tree level
relations between the  masses and mixing angles. The way of
calculating the radiative corrections in the EPA and FDC methods is briefly
described as follows:

In the EPA, the tree level potential $V^{(0)}$ is improved by adding the
1-loop terms~\cite{EPA}:
\begin{eqnarray}
V^{(1)}(Q^2) = V^{(0)}(Q^2) + \Delta V^{(1)}(Q^2)
\label{eq:pot}
\end{eqnarray}
where
\begin{eqnarray}
\Delta V^{(1)}(Q^2) = {1\over 64 \pi^2} \sum_{\stackrel{quarks}{squarks}}
\mathrm{Str} {\cal M}^4 \left(\log{{\cal M}^2\over Q^2} - {3\over 2}\right)
\label{eq:pot1}
\end{eqnarray}
$V^{(0)}(Q^2)$ is the tree level potential evaluated with couplings
renormalized at the scale $Q^2$, and $\mathrm{Str}$ denotes the supertrace
over the third generation of quark and squark fields contributing to the
generalized squared mass matrix ${\cal M}^2$.

The full 1-loop potential $V^{(1)}$ is rediagonalized yielding the 1-loop
corrected physical masses \MH, \Mh and the effective mixing angle \aeff.
Thereby \MA and $\tan\beta$ are usually considered as the free input
parameters.
Explicit formulae for the 1-loop corrected CP-even
Higgs mass matrices can be found in first paper of ref.~\cite{EPA}. The
physical Higgs boson masses and 1-loop mixing angles are used in the Born
formulae for the production cross sections and decay rates.

In many experimental neutral MSSM Higgs searches a simplified version of
the EPA (\epsi) was used in the analysis of the experimental data.
In this approximation only the leading correction coming from the
top and stop quark loops to the Higgs mass matrix is included. In
addition, the effects of the sfermion mixing are neglected. In this case the
corrected Higgs mass matrix has the simple form:
\begin{eqnarray}
{\cal M}_{CP-even} = {\sin2\beta\over 2}\left(
\begin{array}{cc}
\cot\beta M_Z^2 + \tan\beta M_A^2 & - M_Z^2 - M_A^2\\
- M_Z^2 - M_A^2  & \tan\beta M_Z^2 + \cot\beta M_A^2 + \epsilon
\label{eq:emassmat}
\end{array}
\right)
\end{eqnarray}
where $\epsilon$ summarizes the leading 1-loop corrections:
\begin{eqnarray}
\epsilon = {3 G_F m_t^4 \over \sqrt{2} \pi^2 \sin^2\beta}\log
\left({M_{\tilde{t}_1}M_{\tilde{t_2}}\over m_t^2}\right).
\end{eqnarray}
This yields:
\begin{eqnarray}
M_{H,h}^2 = {M_A^2 + M_Z^2 + \epsilon \over 2} \pm
\sqrt{ {(M_A^2 + M_Z^2)^2 + \epsilon^2 \over 4} - M_A^2 M_Z^2\cos^2 2\beta
+{\epsilon \cos 2\beta \over 2}(M_A^2 - M_Z^2) }
\end{eqnarray}
\begin{eqnarray}
\tan\alpha_{eff} = {-(M_A^2 + M_Z^2)\tan\beta \over
 M_Z^2 + M_A^2\tan^2\beta - (1 + \tan^2\beta) M_h^2 }
\label{eq:ealpha}
\end{eqnarray}
The 1-loop corrected Higgs boson masses \MH, \Mh and the effective mixing
angle \aeff thus depend on only one extra SUSY parameter, namely the
product $M_{\tilde{t}_1}M_{\tilde{t}_2}$ of the scalar top mass eigenvalues,
in addition to the tree level parameters \MA and $\tan\beta$.

\begin{figure}[htbp]
\begin{center}
\mbox{
\epsfig{file=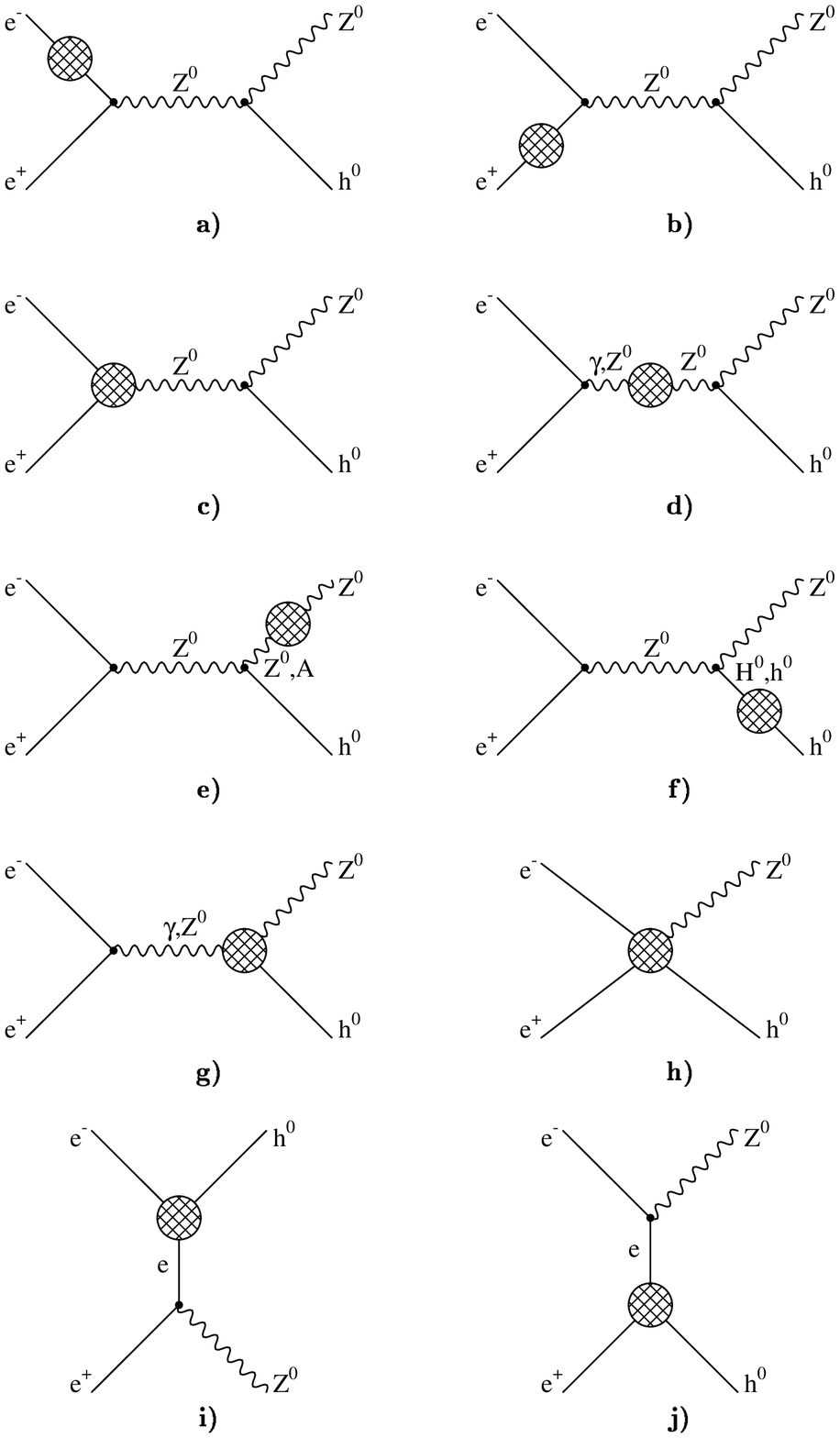,width=0.8\linewidth}}
\end{center}
\caption{
%\small
\baselineskip=12pt
Classes of Feynman diagrams contributing to the
$e^+e^-\ra Z^0h^0$ process in the FDC approach. The  diagrams
contributing to the $e^+e^-\ra A^0h^0$ process can be obtained by changing
$Z^0$ into $A^0$ on the external line and skipping the diagrams i), j).
\label{fig:feyn}
}
\end{figure}

In the FDC approach the 1-loop physical Higgs boson masses are
obtained as the pole positions  of the dressed scalar propagators.
In order to obtain
accurate values of the masses, also the leading reducible terms of higher
order have to be taken into account.
$M_H^2$ and $M_h^2$ are given by
the solution of the equation of the form:
\begin{eqnarray}
\mathrm{Re}\left[\left(p^2 - m_h^2 - \Sigma_{hh}(p^2)\right)
\left(p^2 - m_H^2 - \Sigma_{HH}(p^2)\right) - \Sigma_{hH}^2(p^2) \right]=0
\label{eq:FDCmass}
\end{eqnarray}
For the calculations of the cross sections we need the full set of scalar
2-point functions, and also the 3- and 4-point Green's functions are taken
into account. In Figure~\ref{fig:feyn} the diagrams contributing
to the $e^+e^-\ra Z^0h^0,A^0h^0$ process are collected.
Specially important is the diagram
f) in Figure~\ref{fig:feyn}, illustrating the \hb-\Hb mixing on the external
scalar line, as this mixing corresponds in a good approximation to the effect
of introducing the effective mixing angle \aeff into the Born formulae.
The formulae for the cross sections obtained in the FDC now differ from the
Born expressions, because not only the effective values of the masses are
corrected but also new form factors and momentum dependent effects are
considered. The analytic formulae for the cross sections for
the processes $e^+e^-\ra Z^0h^0, A^0h^0$, including all but the box
and 1-loop $e^+e^-h^0$ vertex contributions,
can be found in the first paper of ref.~\cite{MYNPB}.
Effects of box contributions are discussed in the third paper of
ref.~\cite{MYNPB}. In the present paper we include
also the 1-loop contributions to the $e^+e^-h^0$ vertex
in the cross section calculation.
Note that the diagrams h), i), j) of Figure~\ref{fig:feyn} are not part
of the phenomenological study in~\cite{JRAS}.

\section{Results on Higgs masses and production cross sections}
\label{sec:results}

In this section we present the results for $Z^0h^0$ and $A^0h^0$ production
{}from the complete diagrammatic 1-loop calculation (FDC) and discuss the
quality of simpler approximations:
\begin{itemize}
\item[--] the effective potential approach (EPA), as described above,
\item[--] the simplified EPA with sfermion mixing and bottom/sbottom
loop neglected (\epsi), based on eqs.~(\ref{eq:emassmat}-\ref{eq:ealpha}).
\end{itemize}
In order to give an orientation about the accuracy of the approximations we
show comparisons of the various methods in the different regions of the
parameter space.

\bigskip

{}From the theoretical point of view, the most convenient parameters for the
Higgs sector are the mass \MA of the CP-odd Higgs boson and the ratio
$\tan\beta=\frac{v_2}{v_1}$. From the experimental point of view it is more
natural to use the mass \Mh of the lighter CP-even Higgs instead of
the formal quantity $\tan\beta$. This, however, is technically more
complicated for the calculational scheme. In the first part of this discussion
we follow the conventional way of presentation and choose  $\tan\beta$ as
a free input parameter. In the second part we replace $\tan\beta$ in favor
of the physical neutral Higgs mass \Mh. This has the advantage to directly
get the theoretical results for the production cross sections for the case
that a Higgs boson is found and its mass has been determined.

\begin{figure}[htbp]
\begin{tabular}{p{0.48\linewidth}p{0.48\linewidth}}
\begin{center}
\mbox{
\epsfig{file=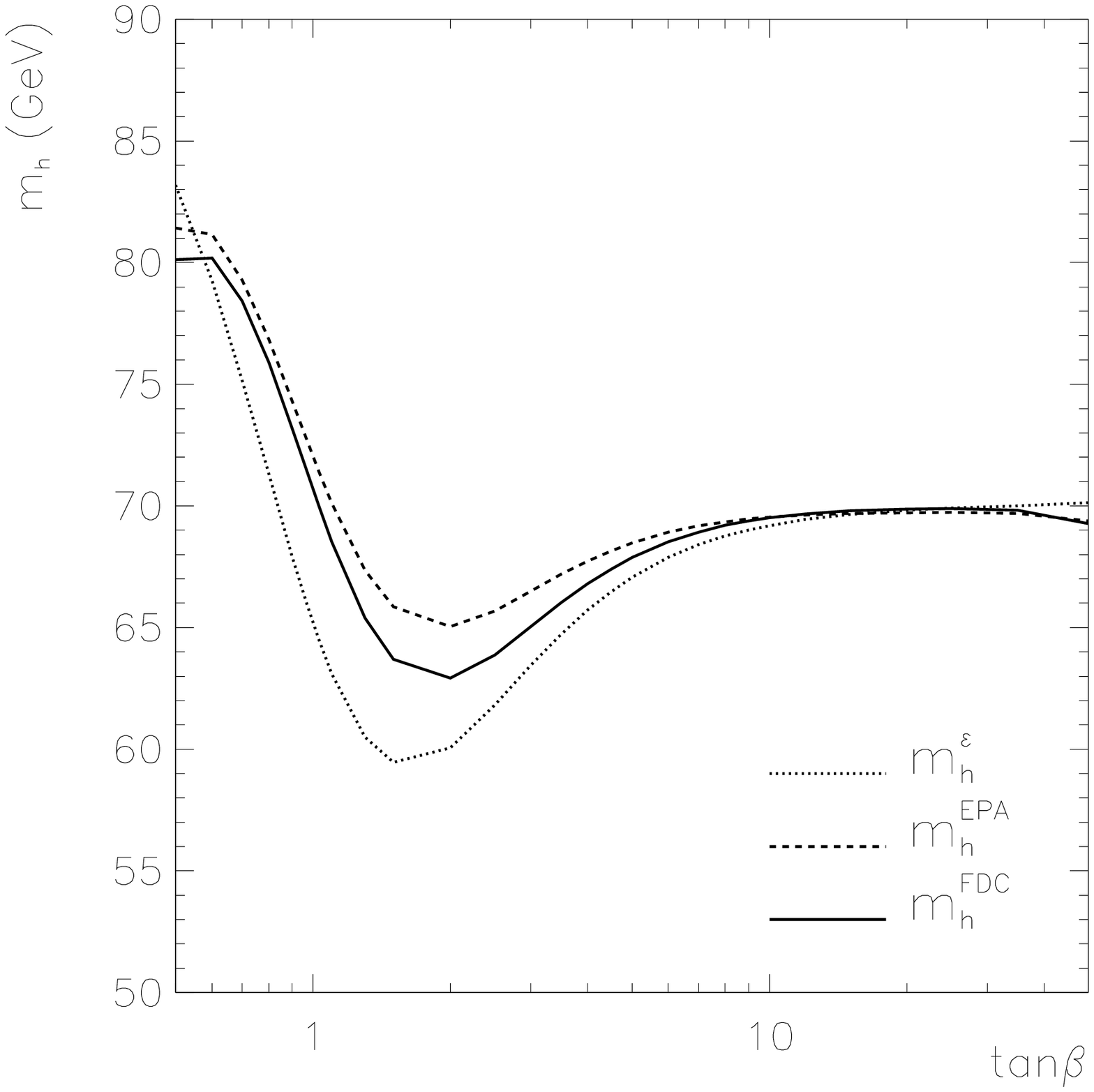,width=\linewidth}}
\end{center}  &
\begin{center}
\mbox{
\epsfig{file=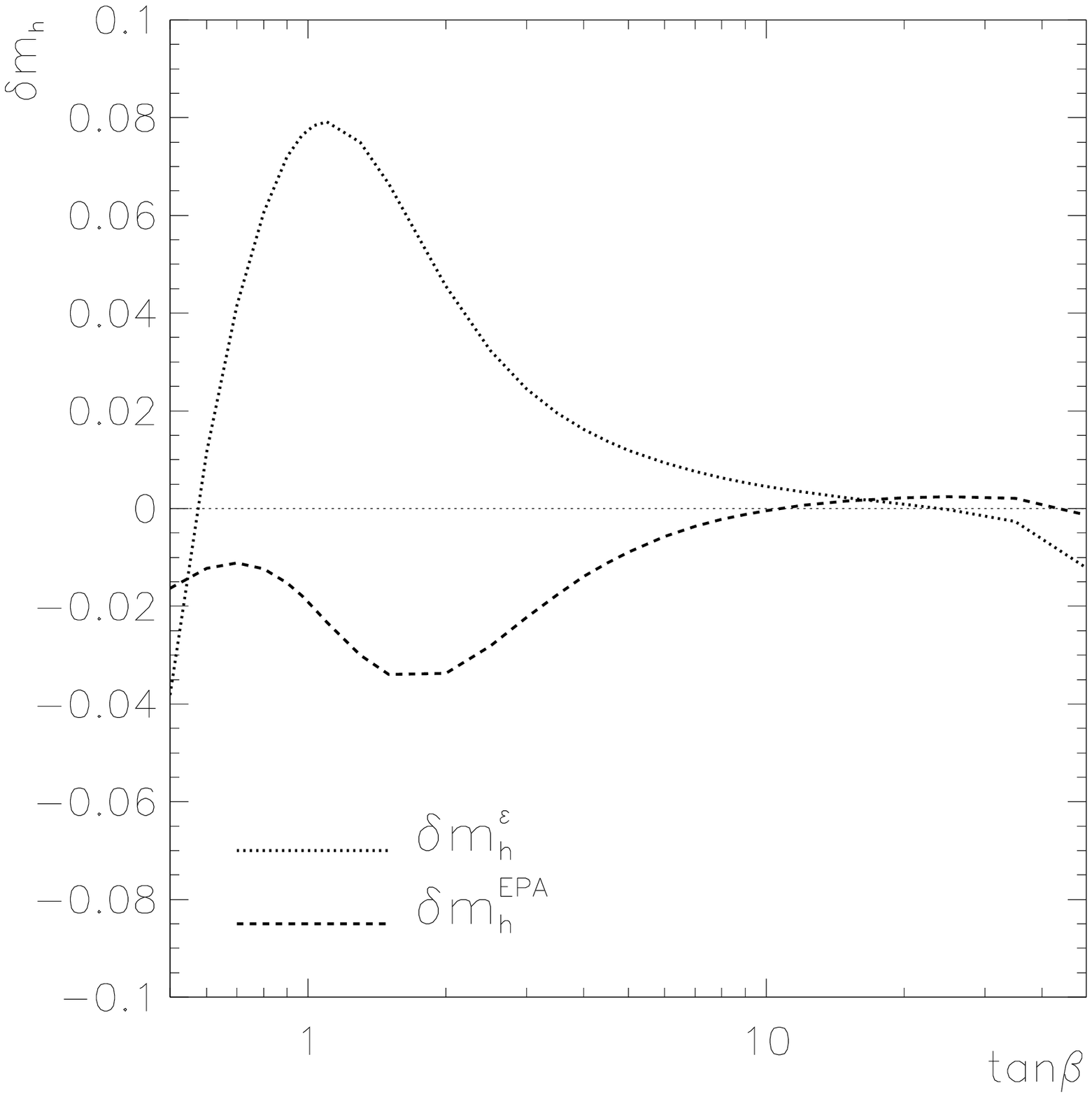,width=\linewidth}}
\end{center} \\
\end{tabular}
\caption{
%\small
\baselineskip=12pt
Comparison of the physical \hb masses obtained in the \epsi, EPA and
FDC. Parameters as given in \protect{Table~\ref{tab:par}}.
\label{fig:mass1}
}
\end{figure}

As a first step, we put together the predictions of the various methods
for the physical \hb mass \Mh for given \MA and $\tan\beta$.
Figure~\ref{fig:mass1} is based on the parameters listed in
Table~\ref{tab:par}.
\begin{table}[htbp]
\begin{center}
\begin{tabular}{|c|c|c|c|c|c|c|c|}
\hline
Parameter & $m_t$ & \MA & $M_{sq}$ & $M_{sl}$ & $M_2$ & $\mu$ & $A_t=A_b$\\
\hline
Value (GeV) & 175 & 70 & 1000 & 300 & 1000 & 500 & 1000\\
\hline
\end{tabular}
\caption{Parameters used for the numerical analysis.
\label{tab:par}}
\end{center}
\end{table}
$\mu$ is the parameter describing the Higgs
doublet mixing in the MSSM superpotential. $M_2$ denotes the SU(2) gaugino
mass parameter. For the U(1) gaugino mass we use the value
$M_1=\frac{5}{3} \tan^2\theta_W M_2$, suggested by GUT constraints.
$\mu$, $M_2$ determine the chargino and neutralino sectors (for the
detailed expressions see for example~\cite{PRD41}).
$M_{sq},M_{sl},A_t$ and $A_b$ are the parameters entering the sfermion mass
matrices. For simplicity we assume a common value $M_{sq}$ for all generations
of squarks, and a common $M_{sl}$ for sleptons.
For the scalar top mass matrix one
has\footnote{Actually in the calculation we use $6\times 6$ sfermion mass
matrices, including the possibility of intergenerational mixing~\cite{PRD41}.
Effects of this mixing appear for example when the mass parameters for
different sfermion generations are not equal to each other.}:
\begin{eqnarray}
{\cal M}_{stop} = \left(
\begin{array}{cc}
(\frac{1}{2} - \frac{2}{3}\sin^2\theta_W) M_Z^2\cos 2\beta
                     + m_t^2 + M_{sq}^2 &
-m_t(\mu\cot\beta + A_t)\\
-m_t(\mu\cot\beta + A_t) &
{2\over 3}\sin^2\theta_WM_Z^2\cos 2\beta + m_t^2 + M_{sq}^2
\end{array}
\right)
\end{eqnarray}

\bigskip

\begin{figure}[htbp]
\begin{tabular}{p{0.48\linewidth}p{0.48\linewidth}}
\begin{center}
\mbox{
\epsfig{file=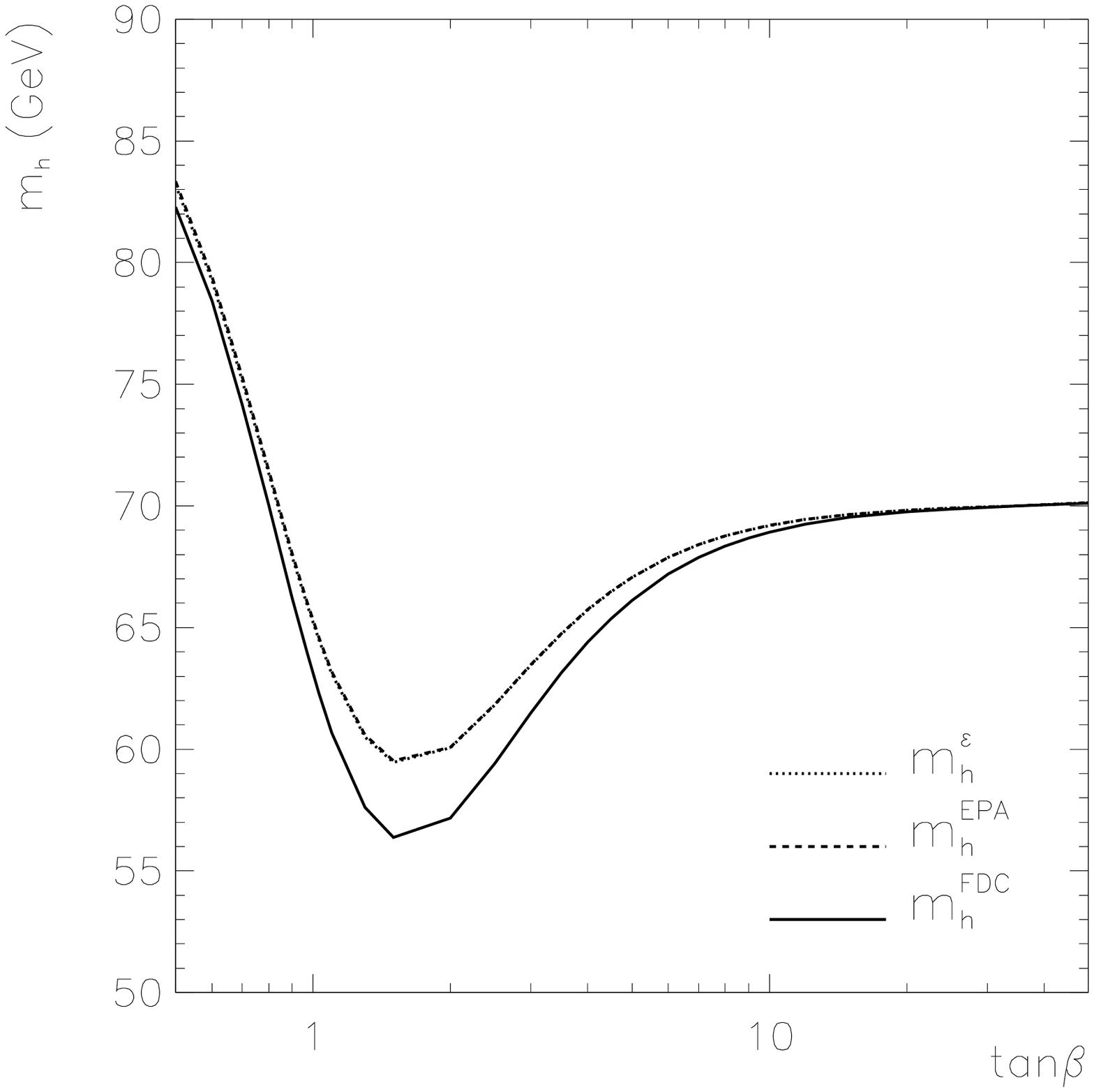,width=\linewidth}}
\end{center}  &
\begin{center}
\mbox{
\epsfig{file=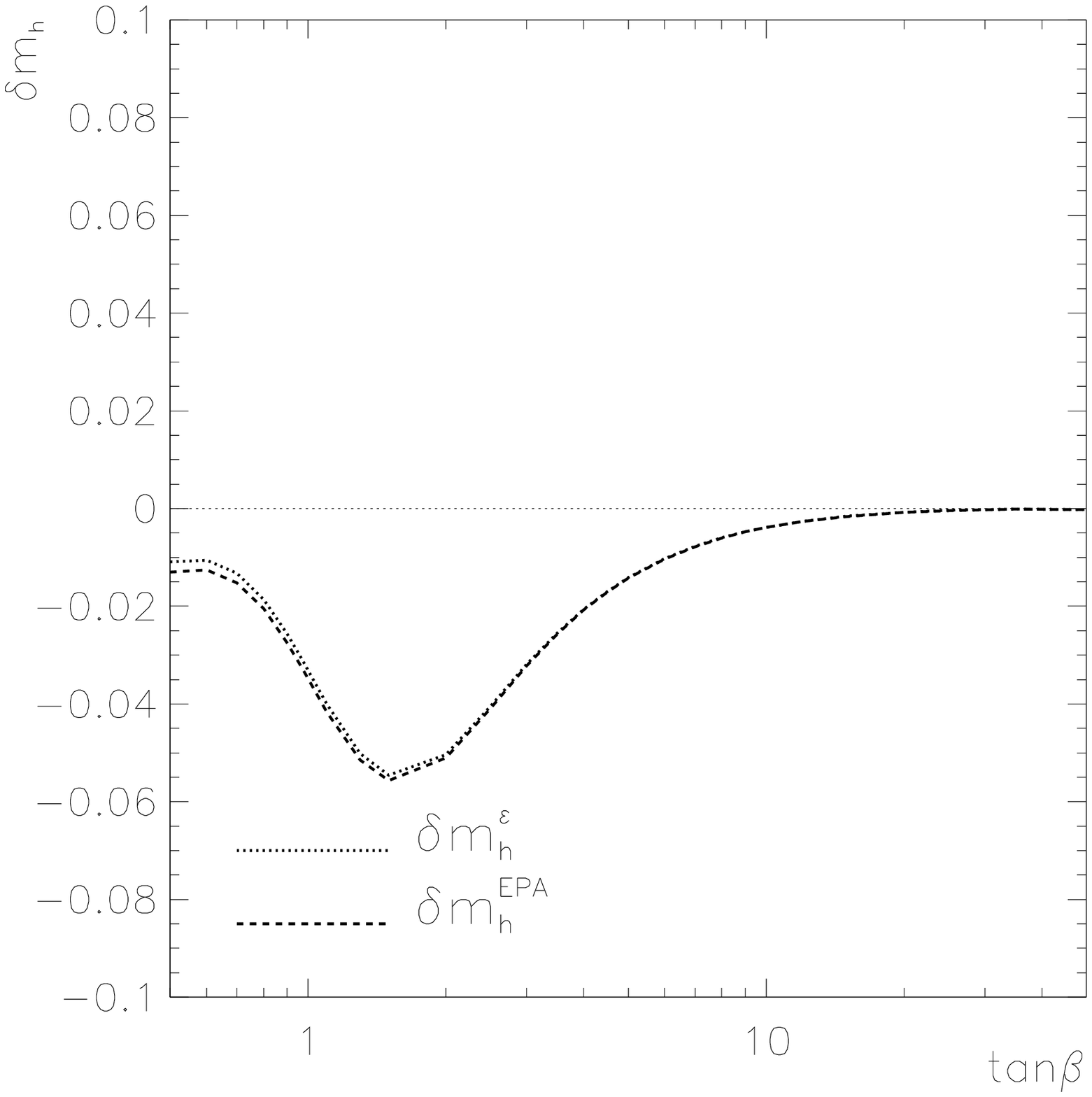,width=\linewidth}}
\end{center} \\
\end{tabular}
\caption{
%\small
\baselineskip=12pt
Comparison of the physical \hb masses obtained in the \epsi, EPA and
FDC. Mass parameters as given in \protect{Table~\ref{tab:par}},
but mixing parameters $\mu,A_t,A_b\approx 0$.
\label{fig:mass0}
}
\end{figure}

In Figure~\ref{fig:mass0},
only the mixing parameters have been turned off, i.e.
$A_t,A_b,\mu\approx 0$ in order to have a situation where the \epsi should be
most suitable. Indeed, the \epsi and the full EPA are almost identical in
this case.

We define the relative differences for the masses with respect to the FDC
as follows:
\begin{eqnarray}
\delta M_h^{EPA,\epsilon} = {M_h^{FDC}-M_h^{EPA,\epsilon}\over M_h^{FDC}}.
\label{eq:diffmass}
\end{eqnarray}
As shown in Figure~\ref{fig:mass1}, the values for \Mh obtained by FDC and EPA
are rather close and differ by less than 4\%. FDC gives lower values of \Mh
than the approximations. This is a general result, valid
for all the parameter choices, and connected with the negative contribution
to \Mh from the gaugino/higgsino sector which is not included in the EPA
calculations.
Figure~\ref{fig:mass1} also shows that the \epsi for \Mh differs more
significantly from the FDC result. The reason is just the neglected mixing
term.

The results for the heavier CP-even scalar mass \MH from FDC and
EPA are again in good agreement. The \epsi for this case yields masses
typically 15\%-20\% different from the FDC values.

\begin{figure}[htbp]
\begin{tabular}{p{0.48\linewidth}p{0.48\linewidth}}
\begin{center}
\mbox{
\epsfig{file=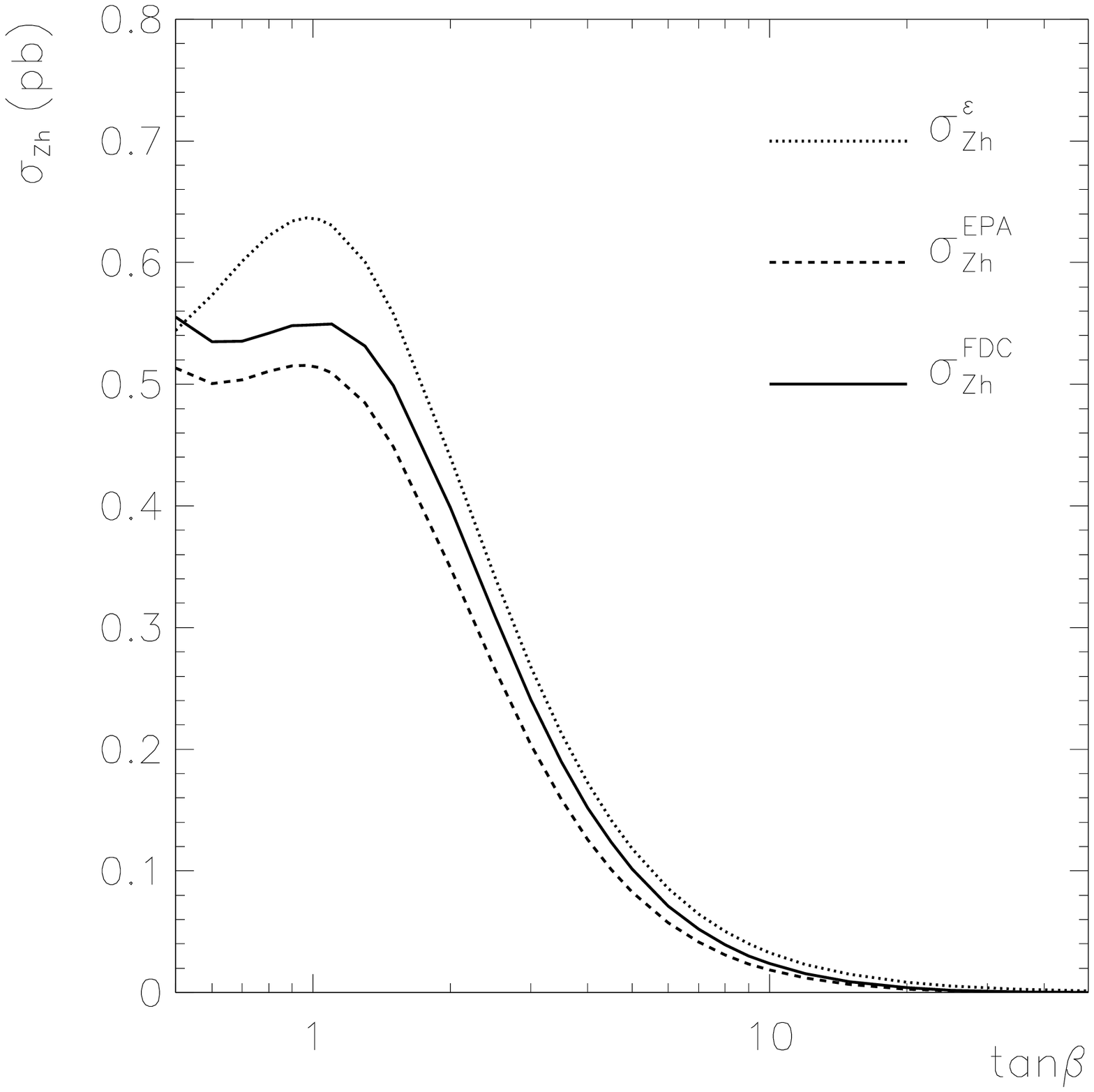,width=\linewidth}}
\end{center}  &
\begin{center}
\mbox{
\epsfig{file=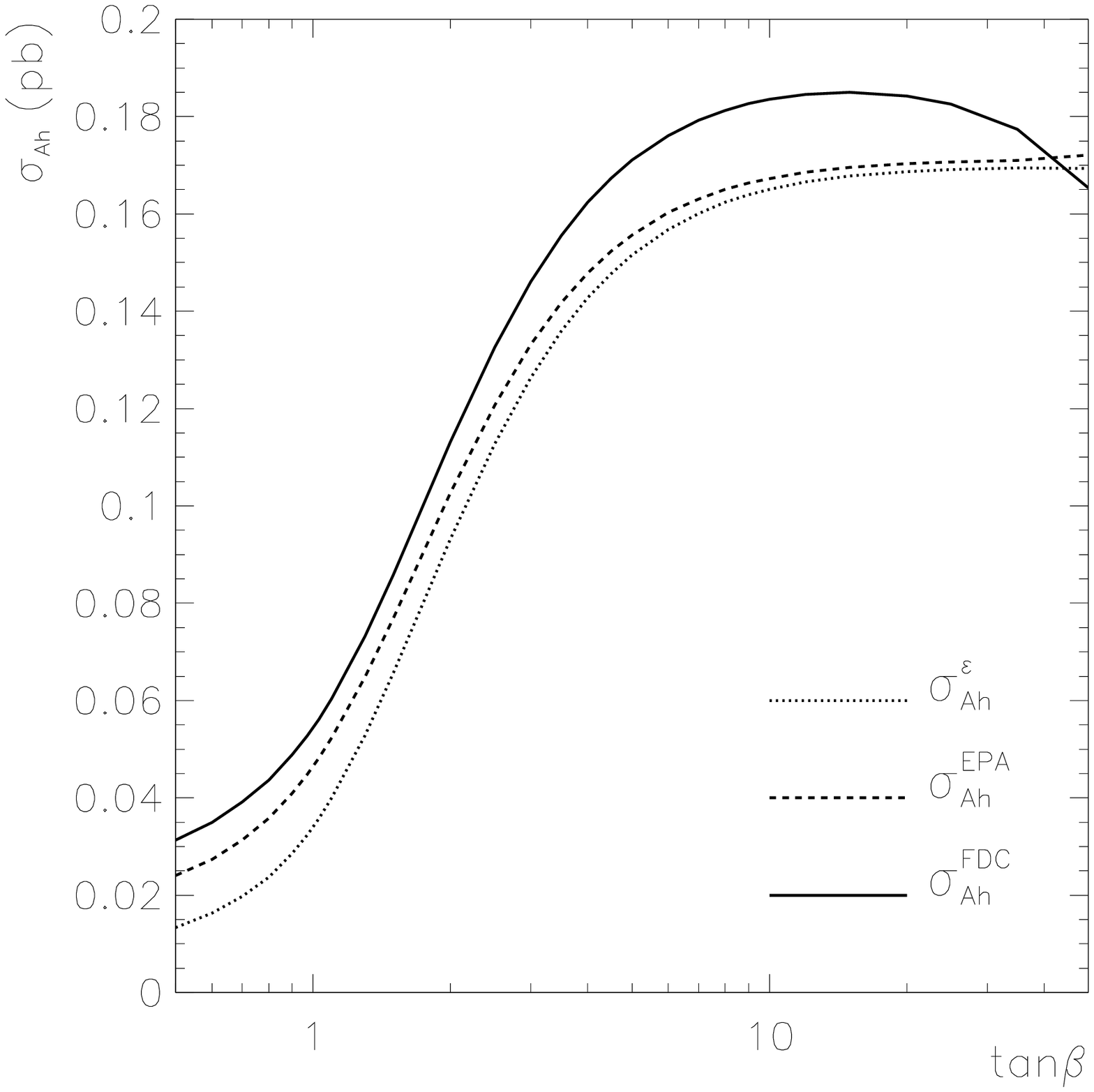,width=\linewidth}}
\end{center} \\
\begin{center}
\mbox{
\epsfig{file=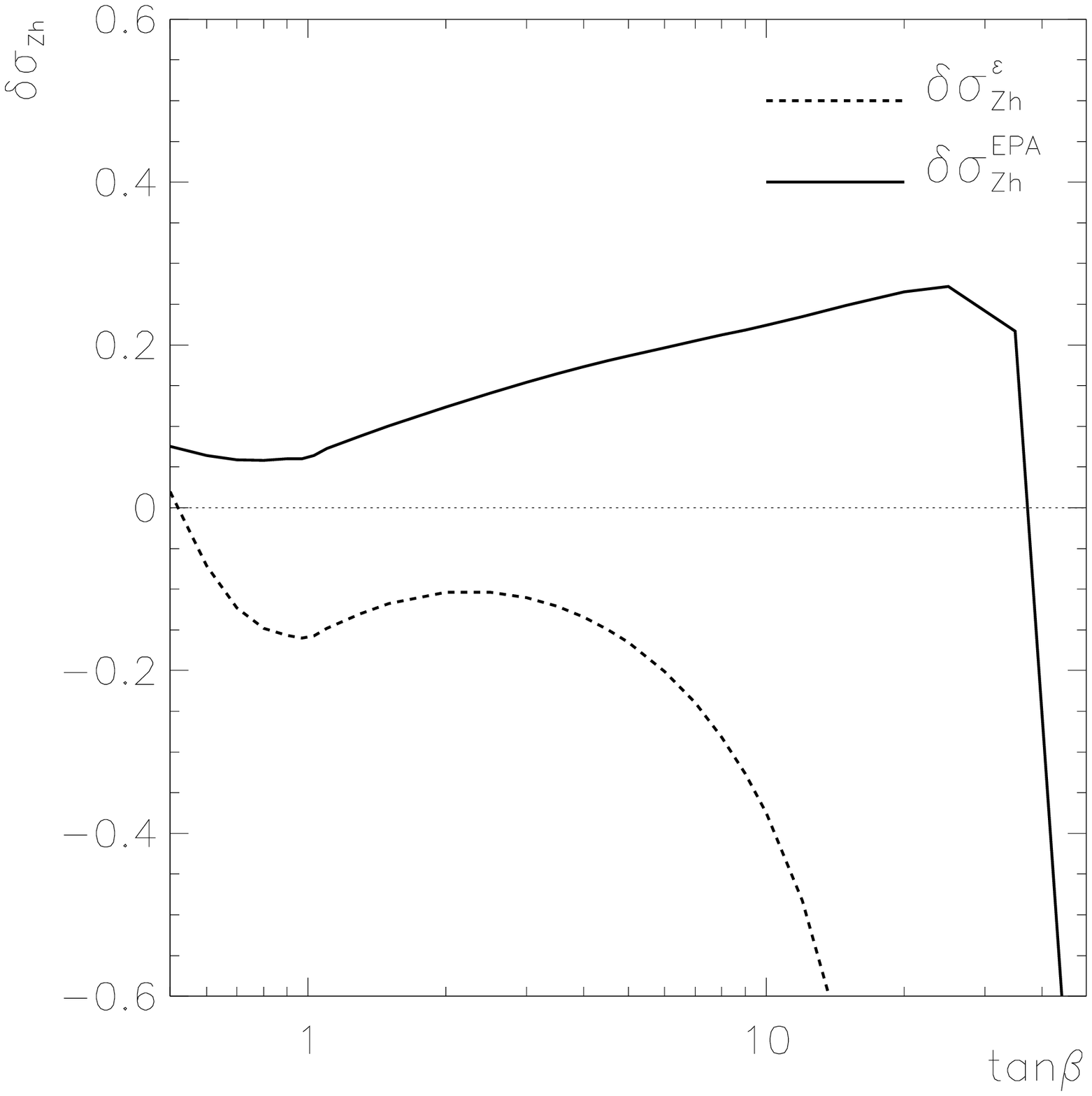,width=\linewidth}}
\end{center}  &
\begin{center}
\mbox{
\epsfig{file=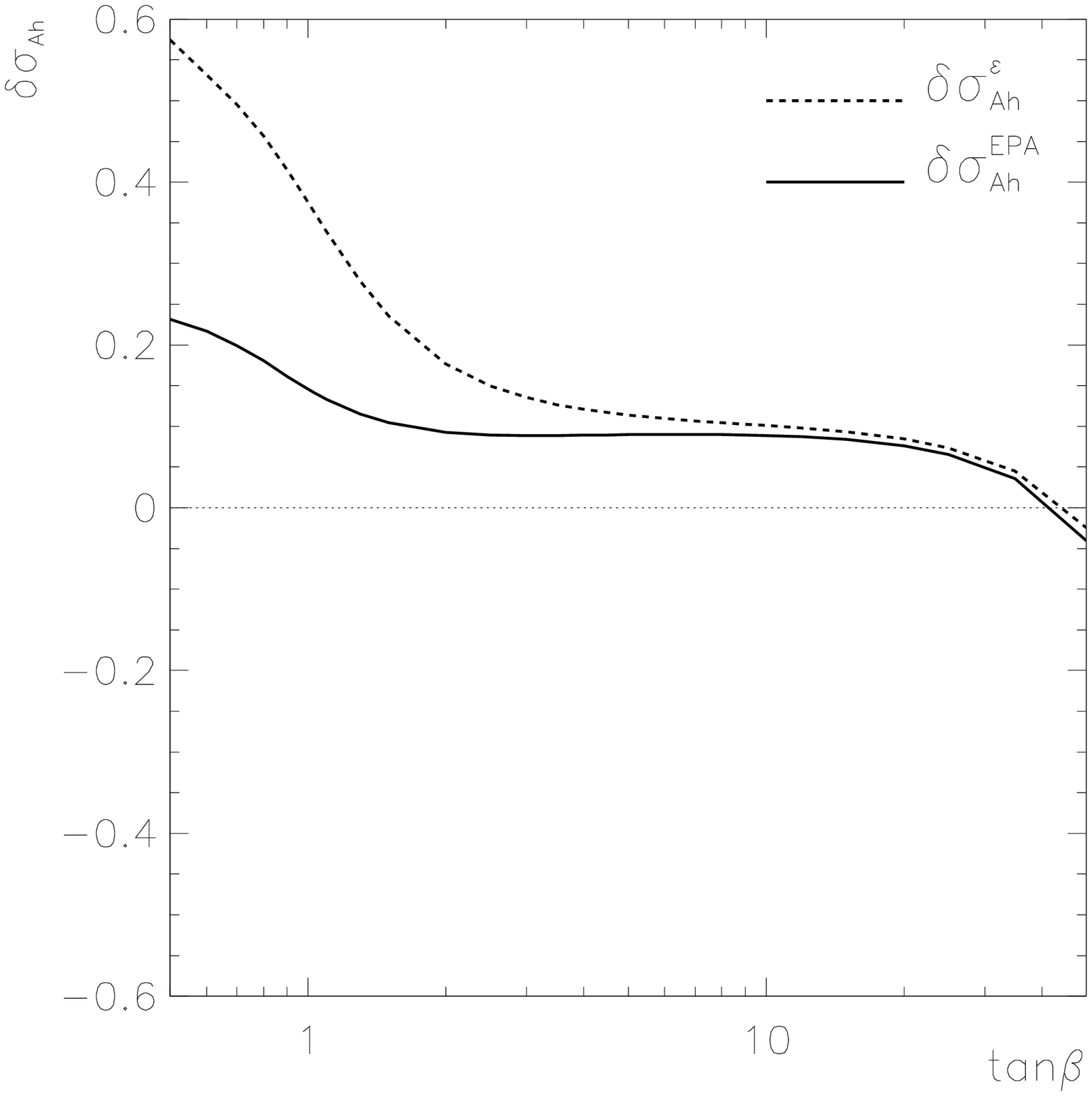,width=\linewidth}}
\end{center} \\
\end{tabular}
\caption{
%\small
\baselineskip=12pt
Comparison of the cross sections \mbox{$\sigma(e^+e^-\ra Z^0h^0,
h^0A^0)$}
obtained in the \epsi, EPA and FDC. Parameters as given in
\protect{Table~\ref{tab:par}}, center-of-mass energy \sqrts = 205 GeV.
\label{fig:cr200}
}
\end{figure}

\begin{figure}[htbp]
\begin{tabular}{p{0.48\linewidth}p{0.48\linewidth}}
\begin{center}
\mbox{
\epsfig{file=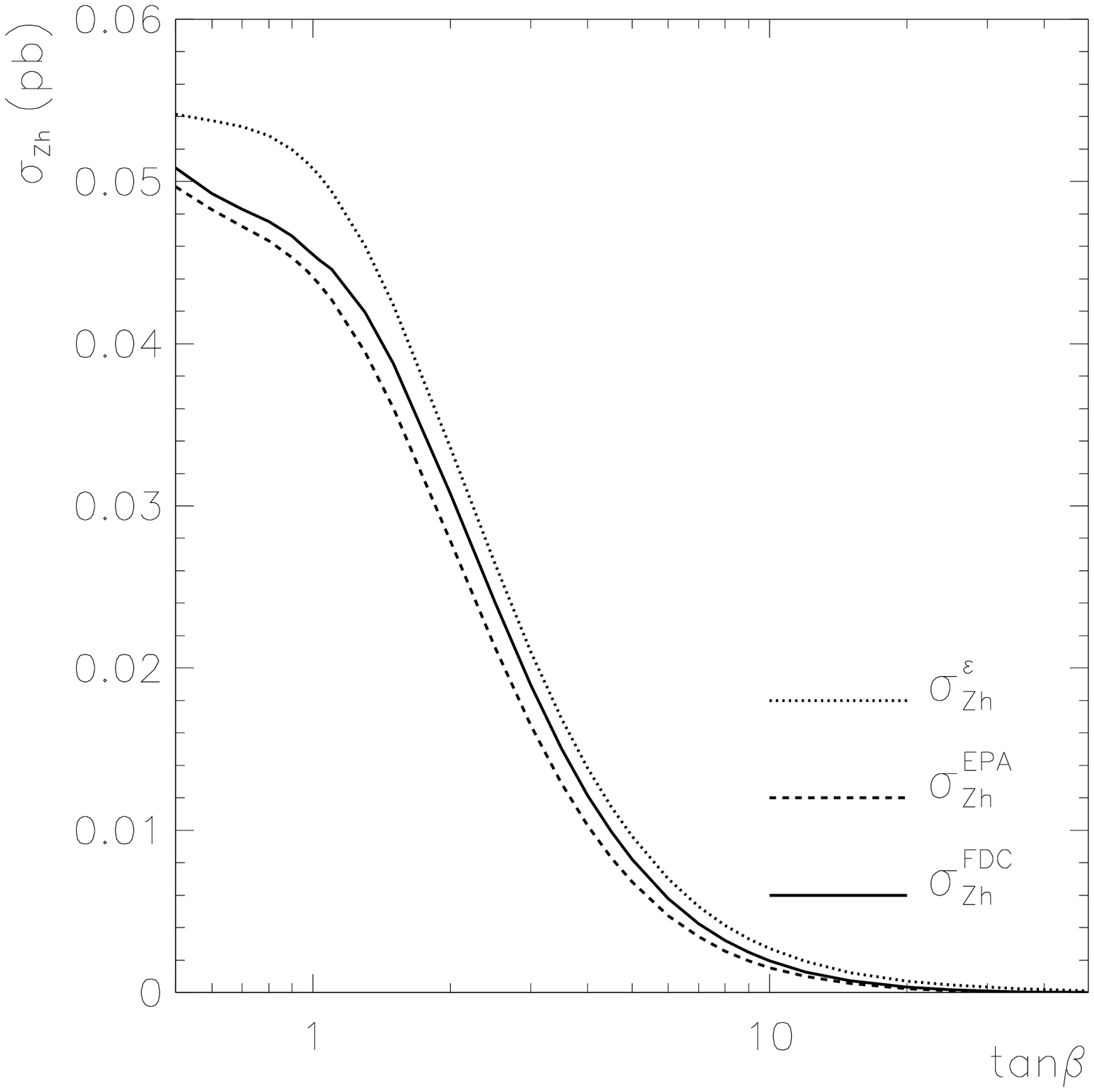,width=\linewidth}}
\end{center}  &
\begin{center}
\mbox{
\epsfig{file=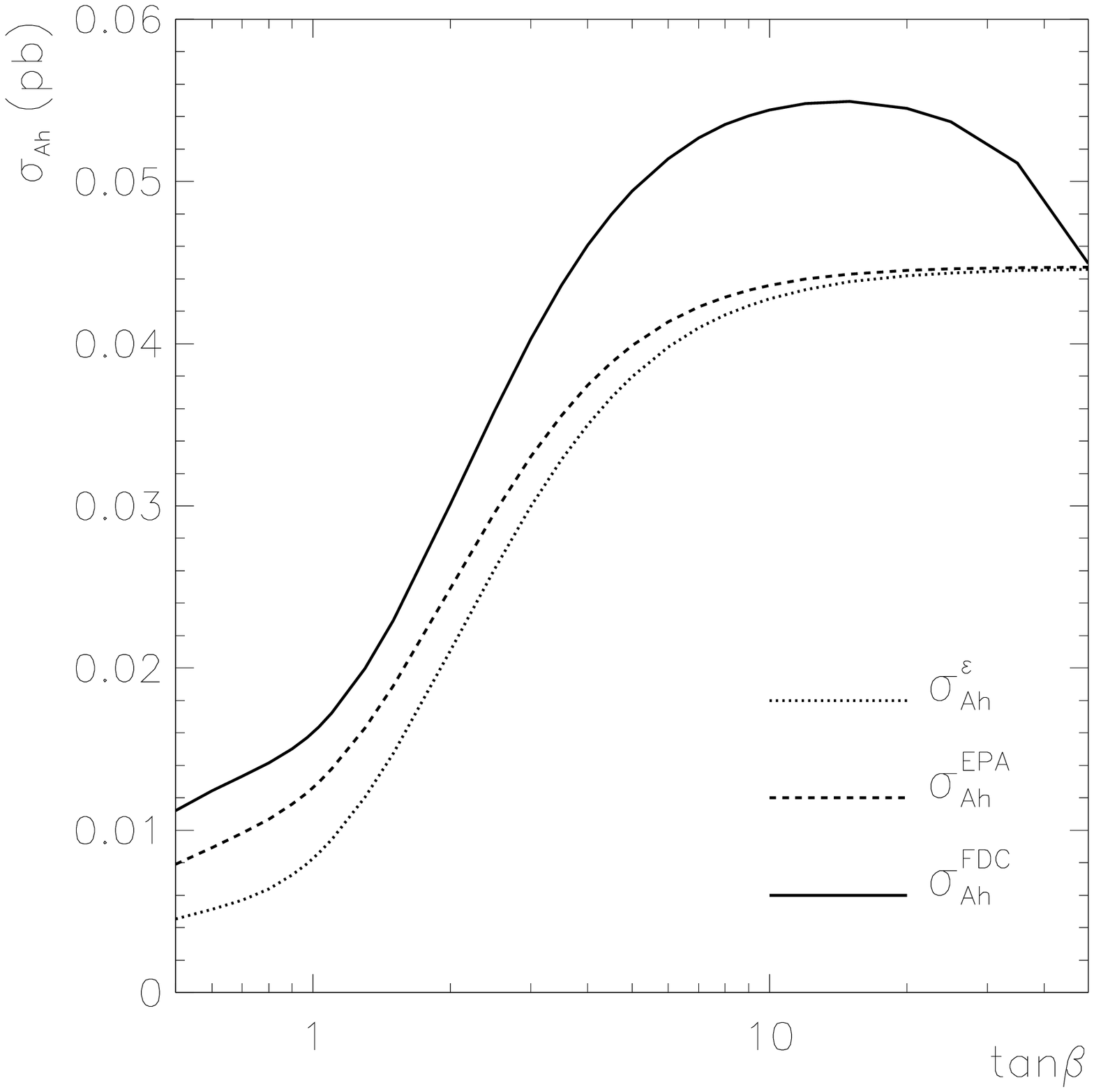,width=\linewidth}}
\end{center} \\
\begin{center}
\mbox{
\epsfig{file=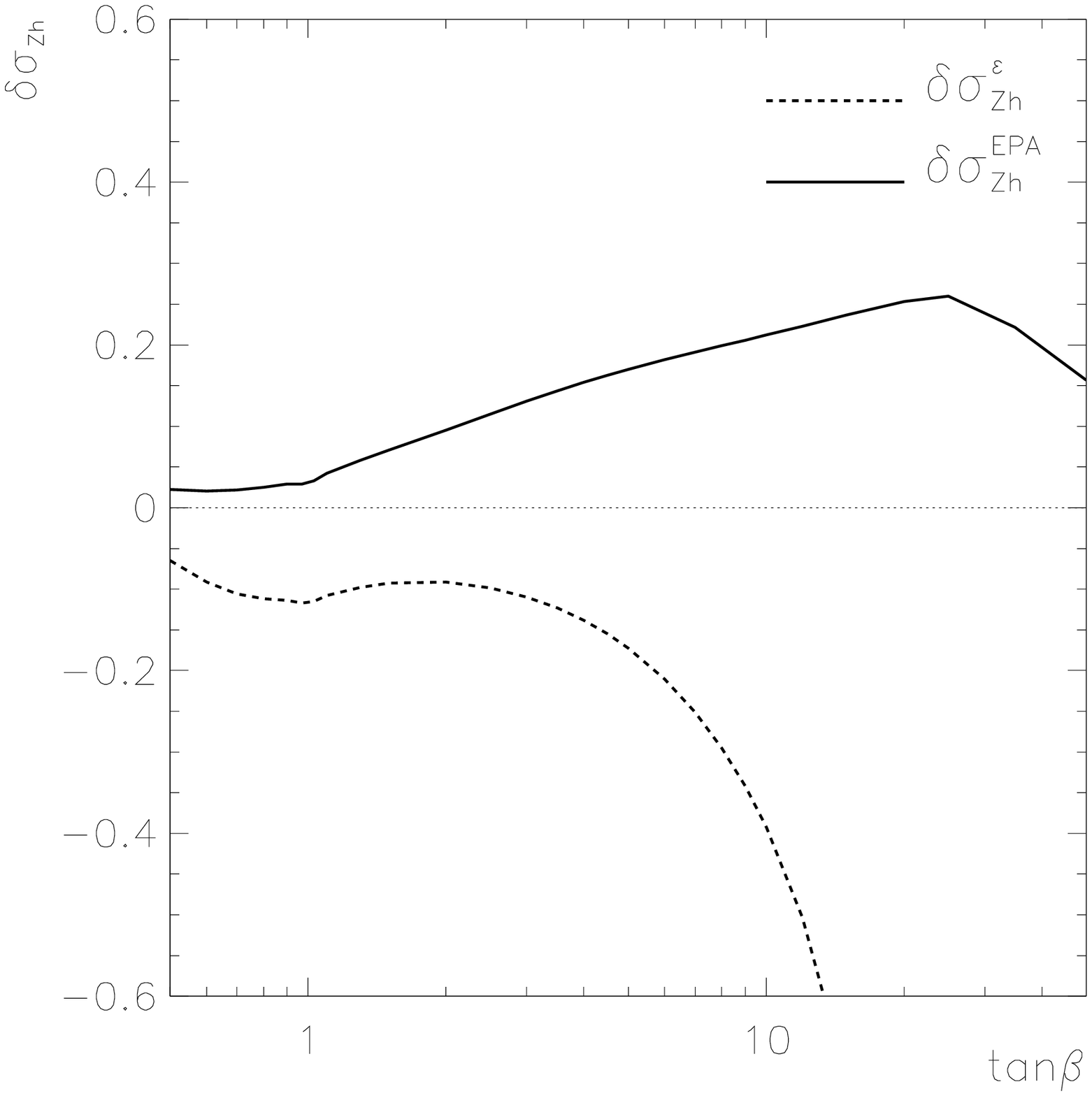,width=\linewidth}}
\end{center}  &
\begin{center}
\mbox{
\epsfig{file=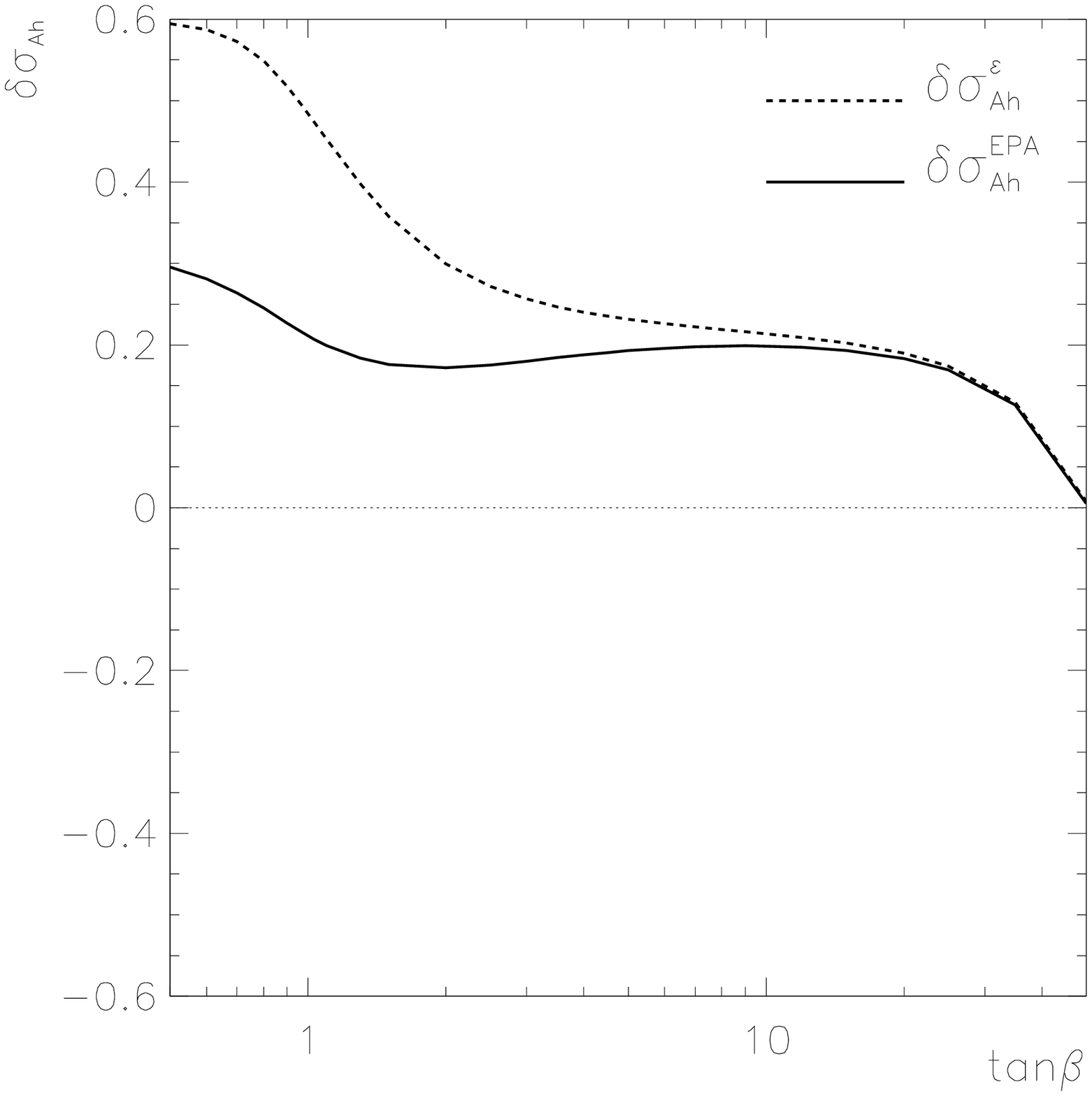,width=\linewidth}}
\end{center} \\
\end{tabular}
\caption{
%\small
\baselineskip=12pt
Comparison of the cross sections \mbox{$\sigma(e^+e^-\ra Z^0h^0,
h^0A^0)$}
obtained in the \epsi, EPA and FDC. Parameters defined as in
\protect{Table~\ref{tab:par}}, center-of-mass energy \sqrts = 500 GeV.
\label{fig:cr500}
}
\end{figure}

Figures~\ref{fig:cr200}, \ref{fig:cr500} display comparisons of
the cross sections for the processes $e^+e^-\ra Z^0h^0, A^0h^0$
for two center-of-mass energies: 205 GeV (Figure~\ref{fig:cr200})
and 500 GeV (Figure~\ref{fig:cr500}). The parameters are the same as
the ones listed in Table~\ref{tab:par}. In analogy to (\ref{eq:diffmass})
we define:
\begin{eqnarray}
\delta\sigma_{Zh}^{EPA,\epsilon}
= {\sigma_{Zh}^{FDC}-\sigma_{Zh}^{EPA,\epsilon}\over \sigma_{Zh}^{FDC}}
{}~~~~~~~~~~~~\sigma_{Zh} = \sigma(e^+e^-\ra Z^0 h^0)\nonumber\\
\delta\sigma_{Ah}^{EPA,\epsilon}
= {\sigma_{Ah}^{FDC}-\sigma_{Ah}^{EPA,\epsilon}\over \sigma_{Ah}^{FDC}}
{}~~~~~~~~~~~~\sigma_{Ah} = \sigma(e^+e^-\ra h^0 A^0).
\label{eq:diffcr}
\end{eqnarray}

As one can see from both figures, the EPA predictions follow in general more
closely the FDC results than the \epsi. The numerical differences can reach
10-20\% at 205 GeV and 30\% at 500 GeV. They become more important with
increasing energies, exceeding 40\% at 1 TeV. Note, however, that in the
region of large cross sections the EPA accuracy is better (20\% up to 500 GeV).

The formally large relative deviations appearing in $\sigma_{Zh}$ for large
$\tan\beta$ are not important from a practical point of view since the cross
section is extremely small in this region.

The less accurate number obtained in the \epsi can again be addressed to the
neglected mixing effects in the scalar quark sector.

To give a more global impression of the typical size of
the differences between the EPA and FDC results,
we have chosen 1000 random points (for each \sqrts value in
Table~\ref{tab:avdif})
{}from the hypercube in the MSSM parameters space with the following bounds:
\begin{center}
\begin{tabular}{lcl}
$0.5<\tb<50$& \mbox{\hskip 1cm }& 5 GeV $<\MA<$ 150 GeV\\
-500 GeV $<\mu<$ 500 GeV&& 200 GeV $<M_2<$ 1000 GeV\\
200 GeV $<M_{sq}<$ 1000 GeV&& 100 GeV $<M_{sl}<$ 300 GeV\\
$-M_{sq}<A_t=A_b<M_{sq}$ & \\
\end{tabular}
\end{center}
We calculated the quantities \admh, \admH, \adszh and \adsah
and averaged them arithmetically
over all generated points of the parameter space.
The results are summarized in Table~\ref{tab:avdif}.
It shows that the predictions of both methods deviate in particular
for $\sigma_{Zh}$.
\begin{table}[htbp]
\begin{center}
\begin{tabular}{|c|c|c|c|}
\hline
Quantity& \sqrts =205 GeV& \sqrts =500 GeV& \sqrts =1 TeV\\
\hline
\admh    & 2.7\% &2.7\%  & 2.7\% \\
\hline
\admH    & 2.2\% & 2.2\% &  2.2\% \\
\hline
$\adszh$ & 22\%  &29\%   & 32\% \\
\hline
$\adsah$ & 5.5\% &9.5\%  & 15\% \\
\hline
\end{tabular}
\caption{Differences between the EPA and FDC predictions averaged over a
random sample of parameters.\label{tab:avdif}}
\end{center}
\end{table}

\begin{figure}[htbp]
\begin{tabular}{p{0.48\linewidth}p{0.48\linewidth}}
\begin{center}
\mbox{
\epsfig{file=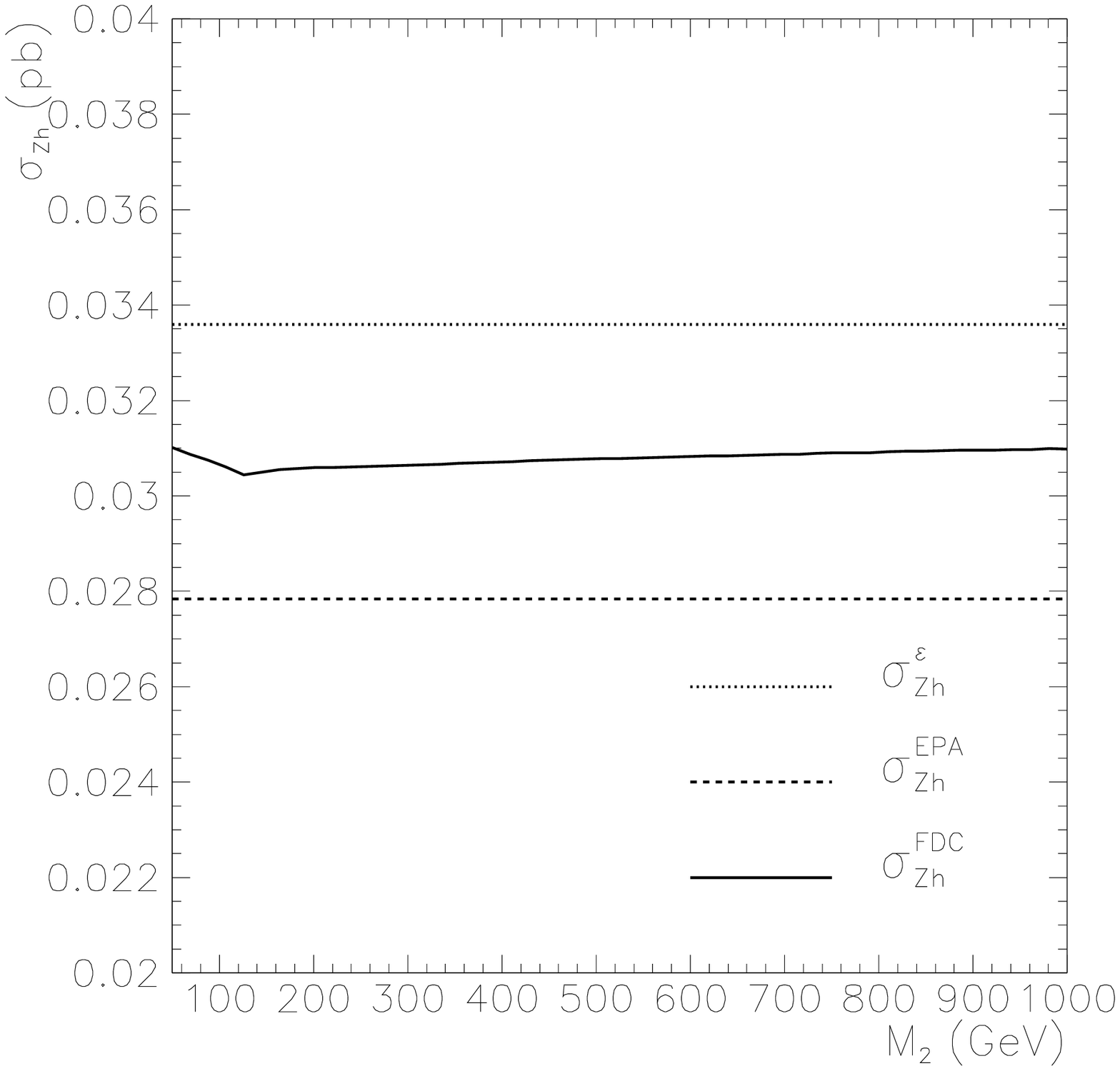,width=\linewidth}}
\end{center}  &
\begin{center}
\mbox{
\epsfig{file=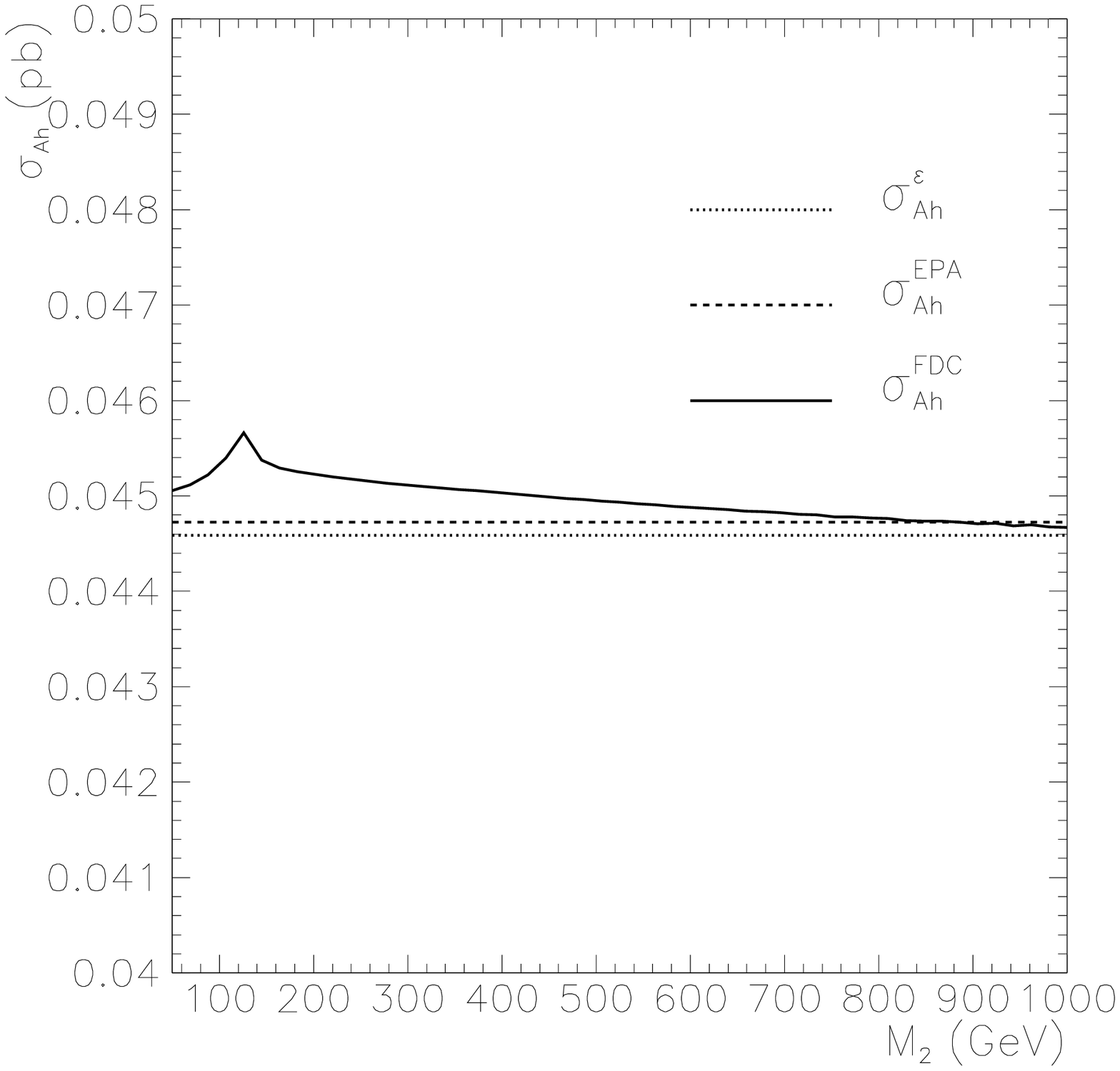,width=\linewidth}}
\end{center} \\
\begin{center}
\mbox{
\epsfig{file=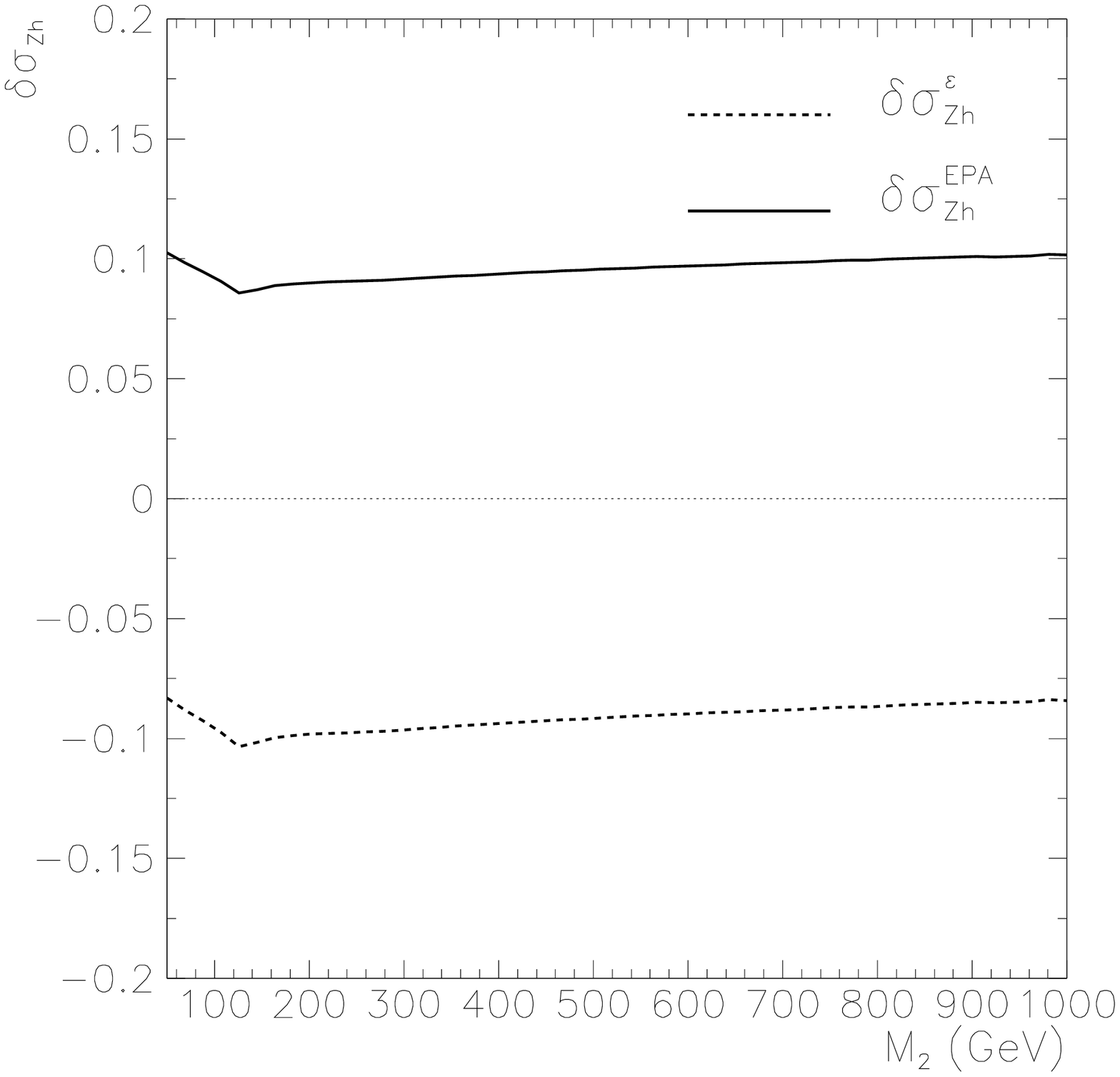,width=\linewidth}}
\end{center}  &
\begin{center}
\mbox{
\epsfig{file=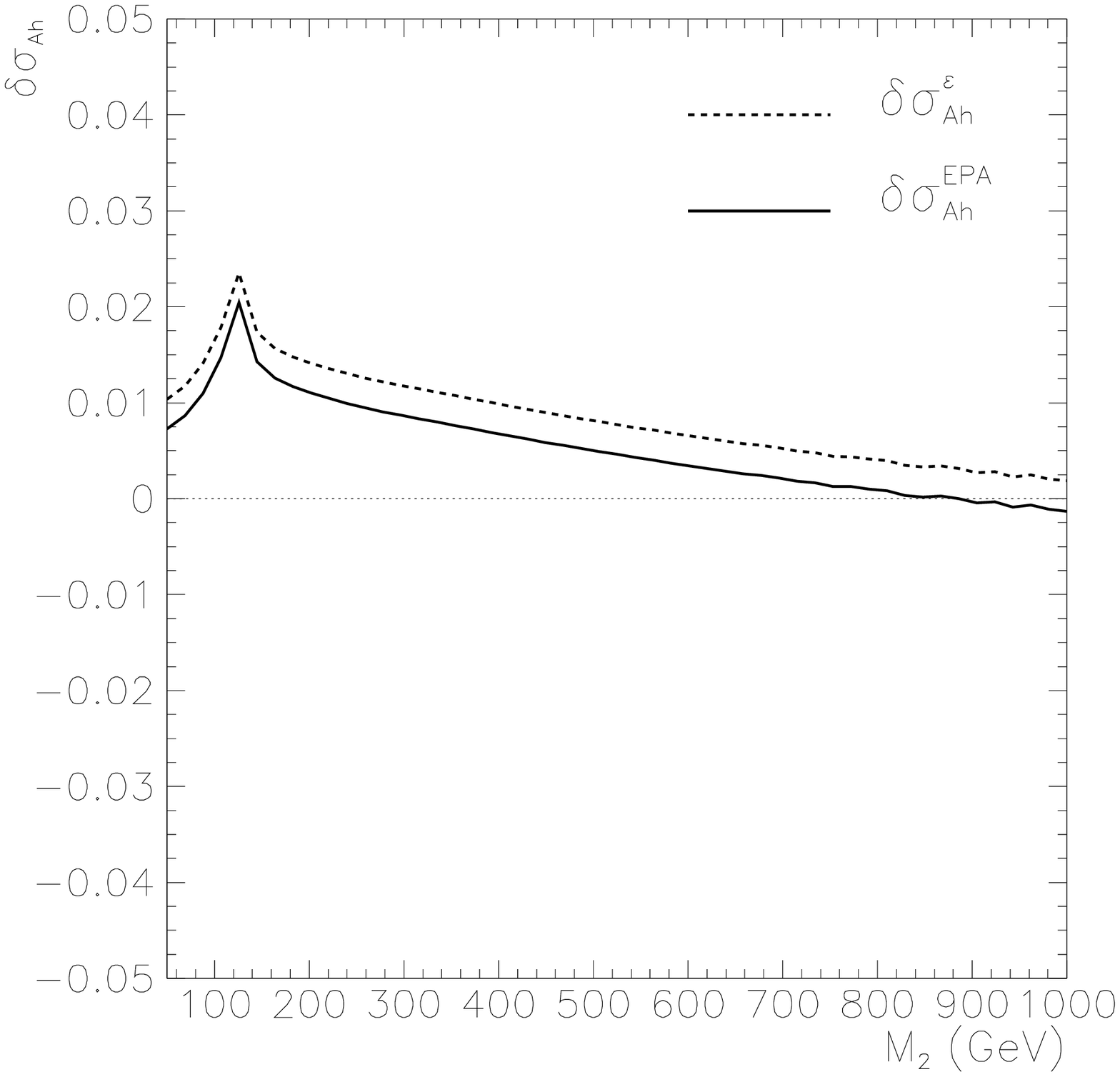,width=\linewidth}}
\end{center} \\
\end{tabular}
\caption{
%\small
\baselineskip=12pt
Comparison of the cross sections \mbox{$\sigma(e^+e^- \ra Z^0 h^0, h^0
A^0)$} as a
function of $M_2$ in the \epsi, EPA and FDC. Parameters as given in
\protect{Table~\ref{tab:par}}. $\tan\beta=2$ in the left plots
and  $\tan\beta=50$ in the right plots.
Center-of-mass energy \sqrts = 500 GeV.
\label{fig:gcr}
}
\end{figure}

We have analyzed also the dependence of the differences between the EPA and
the FDC predictions on the SUSY parameters: sfermion and gaugino masses, $\mu$
parameter and sfermion mixing parameters.
In most cases the variation of those parameters does not
have a large effect on the size of the differences between the EPA and FDC.
The only exception is the gaugino mass, which is completely absent in the EPA.
For a relatively heavy \Ab ($\MA>50$ GeV) the increase of $M_2$ causes
a slow change of \dmh, \dszh, \dsah (see Figure~\ref{fig:gcr}).
This effect can be quite dramatic for small $\MA\sim 5-20$
GeV and $M_2$ around 1 TeV, but such low
% - in some cases $M_h^{FDC}<0$ whereas $M_h^{EPA}>0$,
\MA values are however already excluded by the LEP1 measurements.

\bigskip

We now turn to the more physical parametrization of the cross sections in
terms of the two Higgs boson masses \MA and \Mh. This parameterization is more
clumsy in the calculations, but it has the advantage of physically well defined
input quantities avoiding possible confusions from different renormalization
schemes. Technically, $\tan\beta$ is calculated for given \Mh, \MA and the
other input parameters. Varying \Mh (\MA and other input quantities fixed)
we thus obtain $\tan\beta$ and $\sigma_{Zh}$, $\sigma_{Ah}$ as functions of
\Mh. The predictions for the cross sections are, however, in general not unique
because there can be two solutions for $\tan\beta$, thus yielding two
branches for fixed values of \MA.

\begin{figure}[htbp]
\begin{tabular}{p{0.48\linewidth}p{0.48\linewidth}}
\begin{center}
\mbox{
\epsfig{file=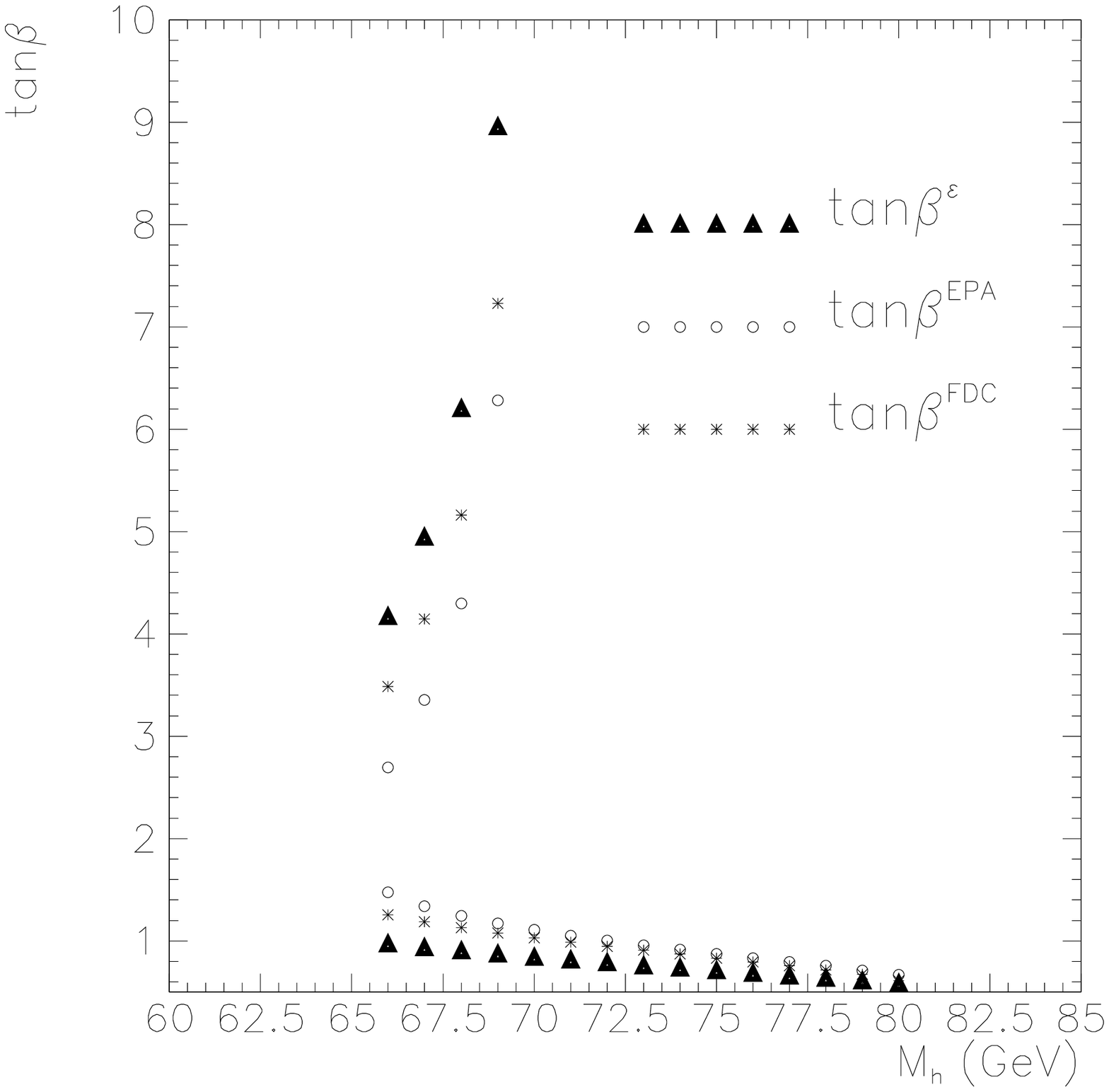,width=\linewidth}}
\end{center}  &
\begin{center}
\mbox{
\epsfig{file=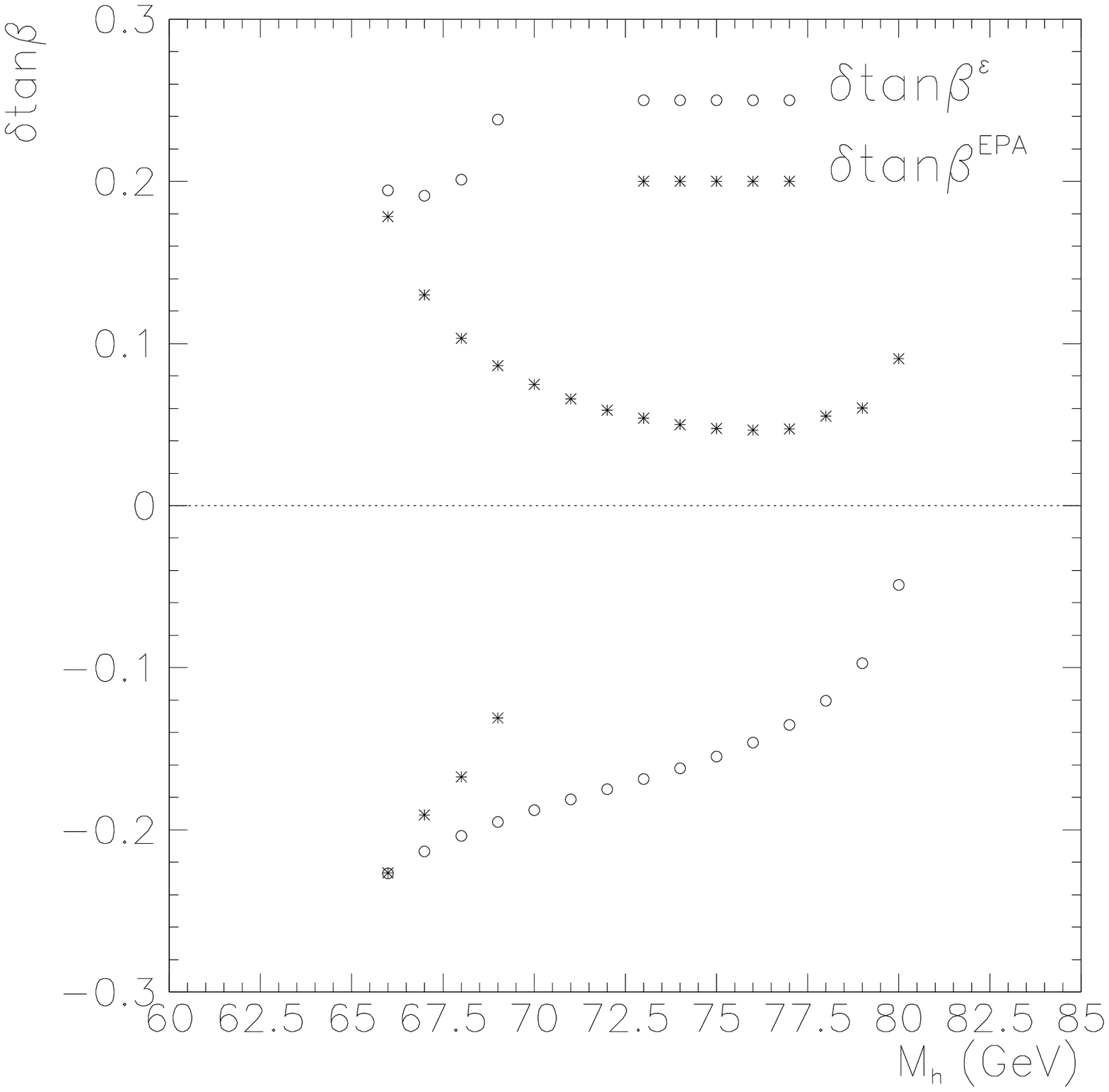,width=\linewidth}}
\end{center} \\
\end{tabular}
\caption{
%\small
\baselineskip=12pt
Comparison of the \tb dependence on \Mh in the \epsi, EPA and
FDC. Parameters as given in \protect{Table~\ref{tab:par}}. $\delta\tan\beta$ is
the relative difference from the FDC values.
\label{fig:tbmass}
}
\end{figure}

\begin{figure}[htbp]
\begin{tabular}{p{0.48\linewidth}p{0.48\linewidth}}
\begin{center}
\mbox{
\epsfig{file=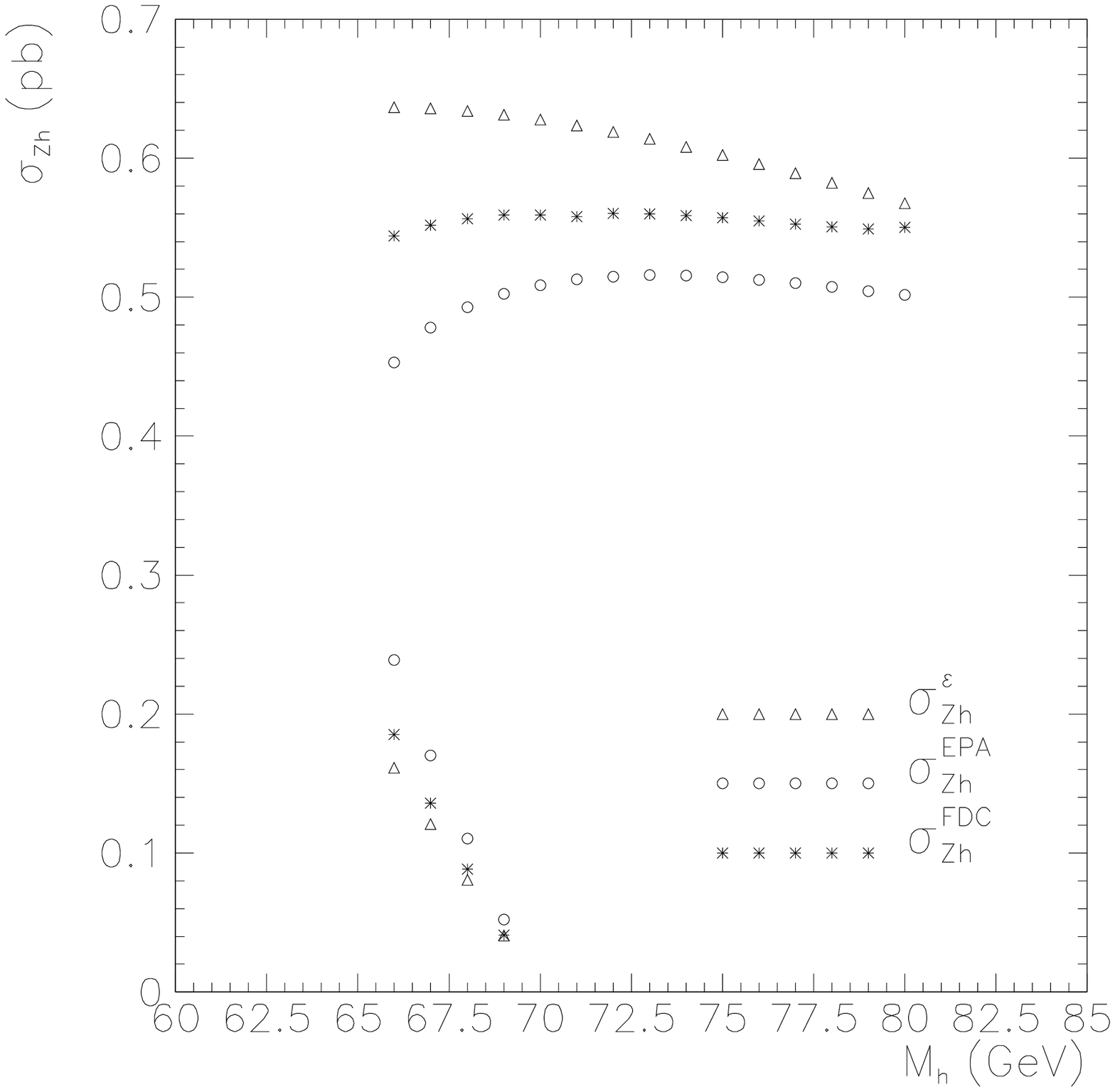,width=\linewidth}}
\end{center}  &
\begin{center}
\mbox{
\epsfig{file=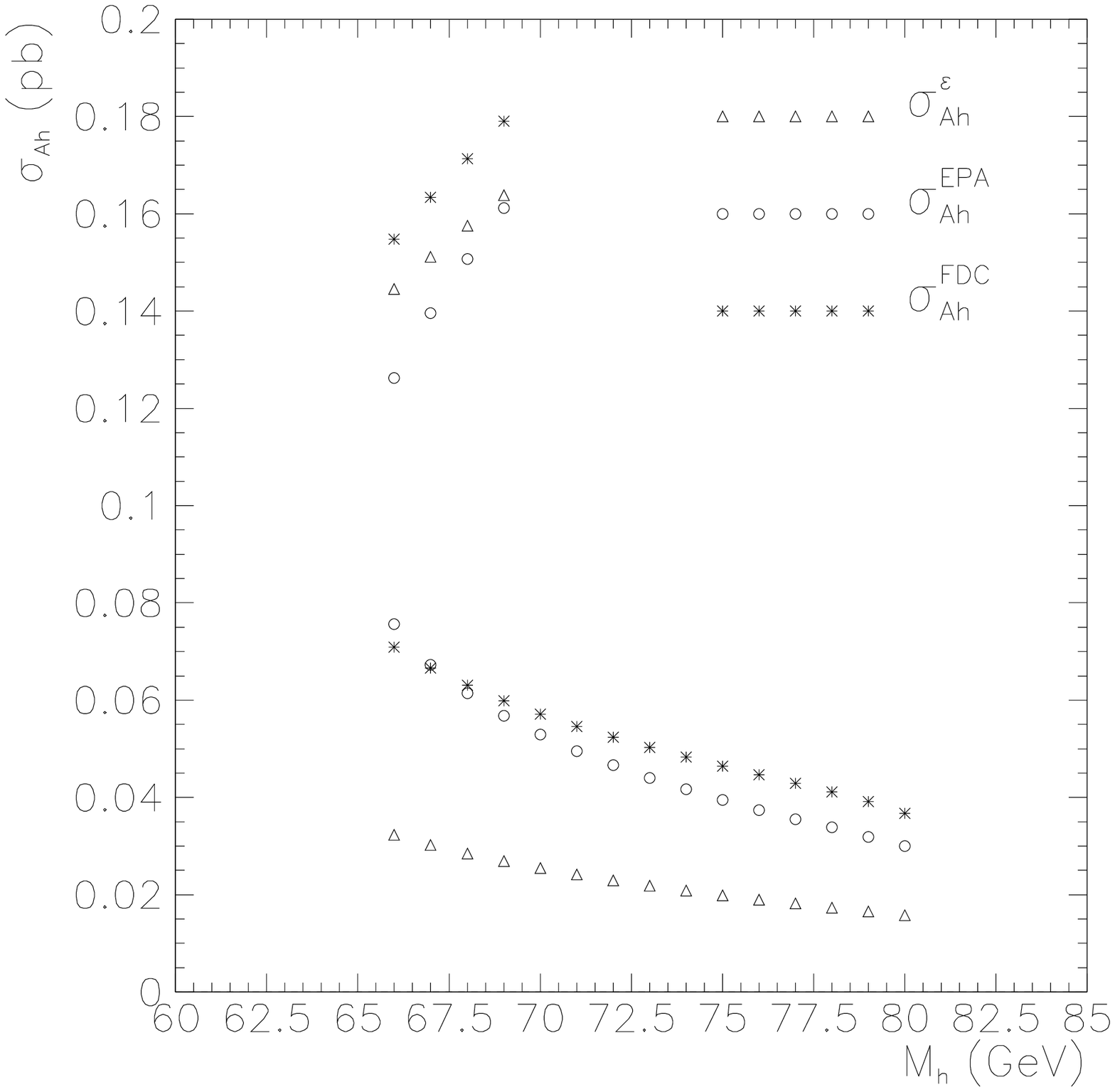,width=\linewidth}}
\end{center} \\
\begin{center}
\mbox{
\epsfig{file=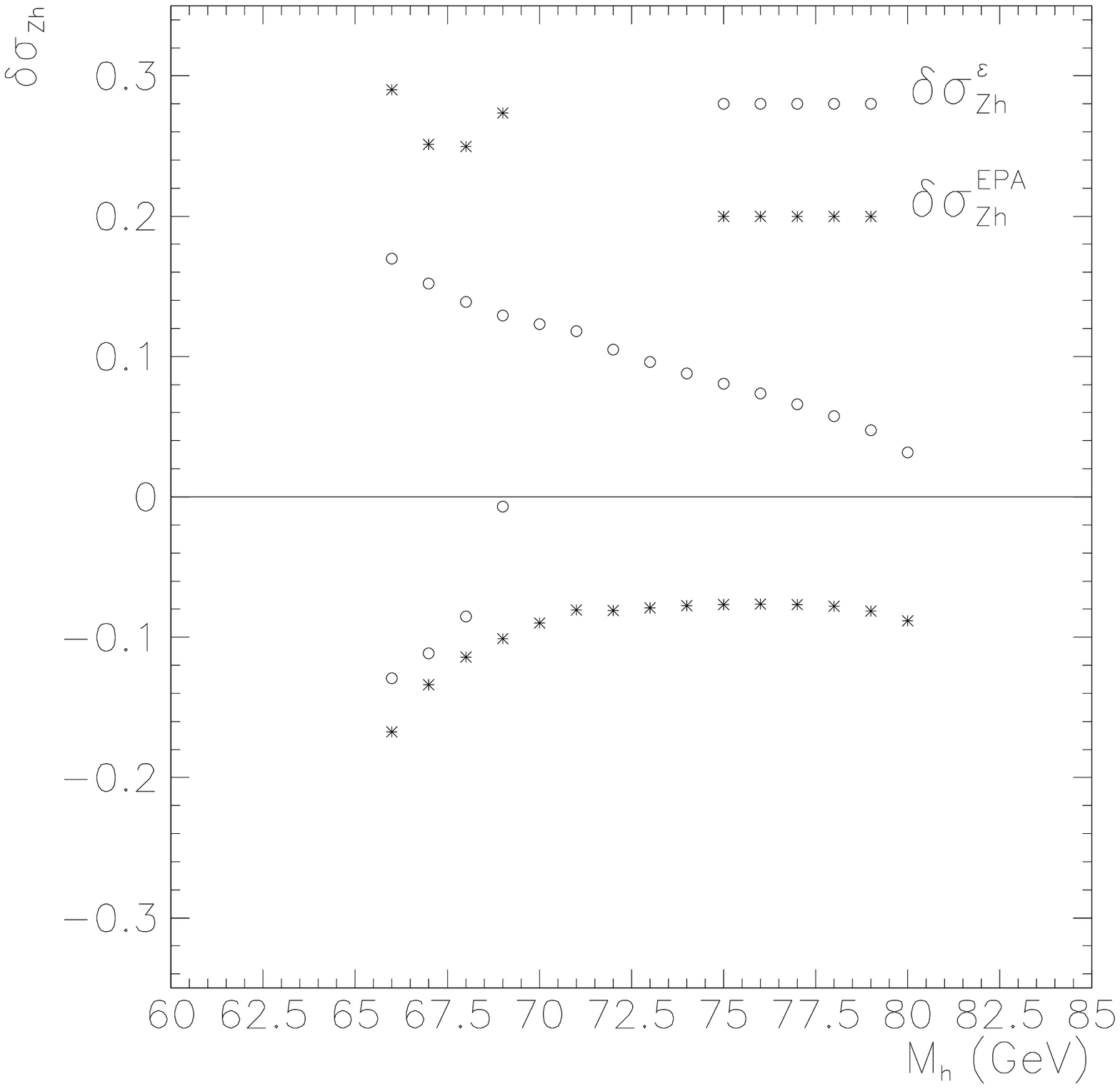,width=\linewidth}}
\end{center}  &
\begin{center}
\mbox{
\epsfig{file=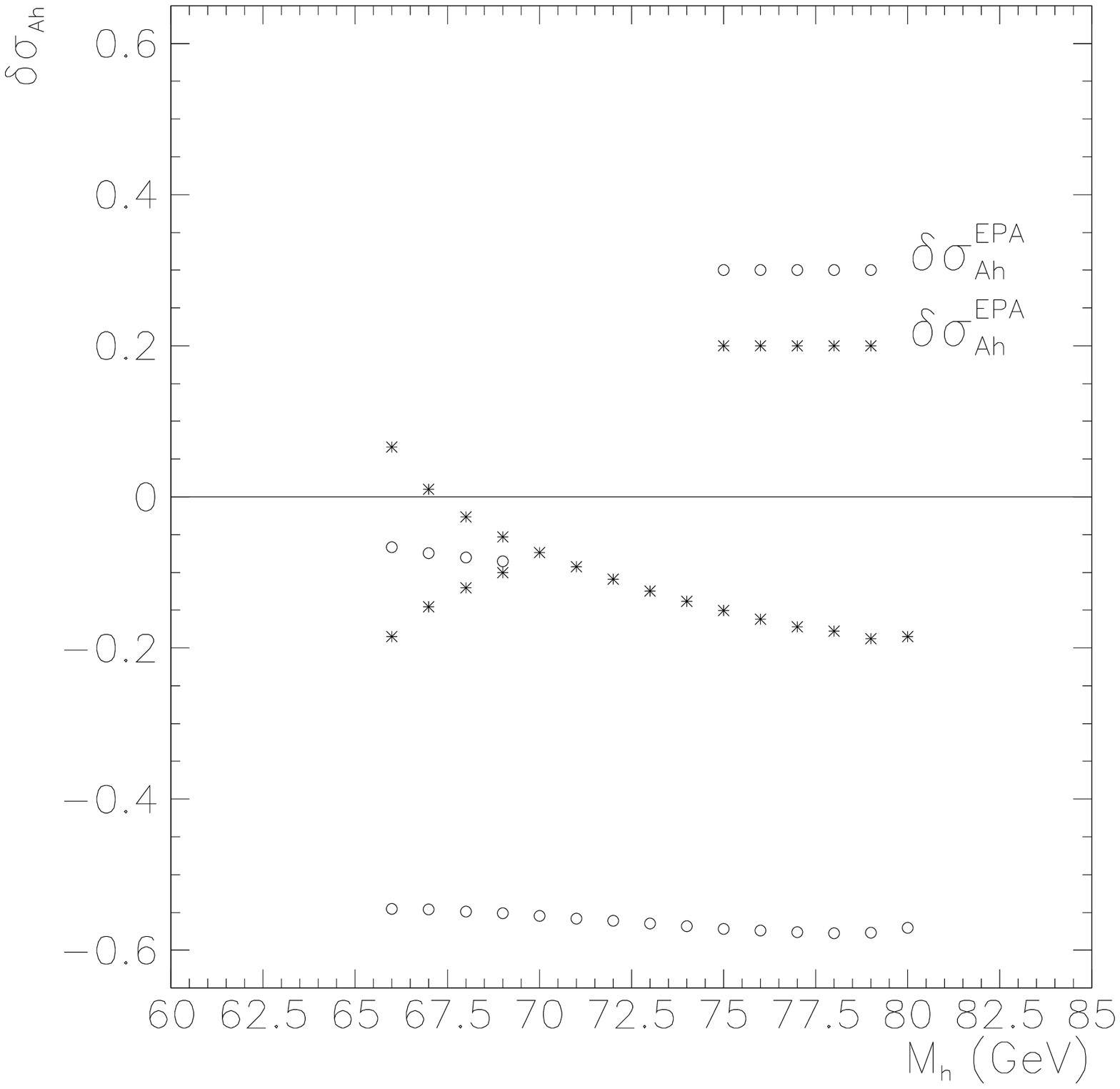,width=\linewidth}}
\end{center} \\
\end{tabular}
\caption{
%\small
\baselineskip=12pt
Comparison of the cross sections \mbox{$\sigma(e^+e^- \ra Z^0 h^0, h^0
A^0)$} as a
function of \Mh in the \epsi, EPA and FDC. Parameters as given in
\protect{Table~\ref{tab:par}}, center-of-mass energy \sqrts = 205 GeV.
\label{fig:tbcr}
}
\end{figure}

The solutions for $\tan\beta$ are shown in Figure~\ref{fig:tbmass} for
the input parameters of Table~\ref{tab:par}, again for the FDC and the simpler
approximations. As can be seen from the figure, the differences between
the \tb values obtained in the EPA and FDC can reach 20\%.
Also significant differences can occur for the cross sections, as displayed
in Figure~\ref{fig:tbcr} where the predictions of EPA and FDC for the
$\sigma_{Zh}$ and $\sigma_{Ah}$ are plotted as functions of \Mh.
The typical size of differences between the methods is
10-20\% for \sqrts =205 GeV.

\section{Summary}
\label{sec:summary}

We presented the results from a complete 1-loop diagrammatic
calculation in the MSSM for the cross
sections for neutral Higgs production via
$e^+e^-\ra Z^0h^0$ and $e^+e^-\ra A^0h^0$, including box diagrams and
$\hb e^+e^-$ vertex contributions which were not part of previous studies
for Higgs searches in the literature.
Comparisons between the FDC predictions with the simpler EPA approximation
have shown that the EPA has an accuracy of typically 10-20\% in the
parameter regions where the cross sections are large.
The differences become bigger with increasing energy. They may
be specially important for future high energy $e^+e^-$ colliders.
The use of the physical input variables $M_A,$ $M_h$ avoids ambiguities
from the definition of $\tan\beta$ in higher order, but the observed
differences remain of the same size.
 For a better accuracy, the full FDC would be
required. It will also be necessary to incorporate the leading
2nd order terms since they are at least as big as the non-leading
1-loop contributions.\\

\noindent
The library of  FORTRAN codes for the calculation of the 1-loop radiative
corrections in the
on-shell renormalization scheme to the MSSM neutral Higgs
production and decay rates~\cite{MYNPB} can be found at the URL address\\
{\sl http://itpaxp1.physik.uni-karlsruhe.de/$\sim$rosiek/neutral\_higgs.html}

\vskip 1cm
\noindent
{\bf Acknowledgments}
\vskip 0.5cm

\noindent
This work was supported in part by the Alexander von Humboldt Stiftung, by the
European Union under contract CHRX-CT92-0004 and by the
Polish Committee for Scientific Research.

\newcommand{\group}[2]{#1 Collaboration, #2 \etal }
\newcommand{\us   }   {\group{L3}{B.~Adeva}}
\newcommand{\usnew}   {\group{L3}{O.~Adriani}}
\newcommand{\etal }   {{\em et~al.}}

\newpage

\end{document}